\documentclass[11pt]{article}
\pdfoutput=1
\usepackage{jheppub,amsmath,amssymb}
\usepackage{mathtools}
\usepackage{amstext}
\usepackage{graphicx}
\usepackage{placeins}
\usepackage{epsfig}
\usepackage{verbatim} 
\usepackage{caption}
\usepackage{slashed}
\usepackage{color}

\usepackage{graphicx}
\usepackage{wrapfig}
\usepackage{upgreek}
\usepackage{bm} %--> bold math...
\usepackage{framed}
\usepackage{bbm}
\usepackage{textcomp}
\usepackage{tikz}
\usepackage{pifont}
\usetikzlibrary{matrix,shapes,fit,tikzmark,calc}
\usepackage{adjustbox}
\usepackage{makecell}
\usepackage{bbold}
\usepackage{tcolorbox}
\usepackage{physics}
\usepackage{empheq}
\usepackage[normalem]{ulem}
\usepackage{subcaption}
\def\bea{\begin{eqnarray}}
\def\eea{\end{eqnarray}}
\def\be{\begin{equation}}
\def\ee{\end{equation}}
\def\ba{\begin{array}}
\def\ea{\end{array}}

\newcommand{\MP}{M_{\rm pl}}

\def\d{{ d}}

\def\0{{\boldsymbol 0}}

\newcommand{\wid}{\text{width}}

\begin{document}
\title{Causality Bounds on the Primordial Power Spectrum}
	\author[a]{Mariana Carrillo Gonz\'alez}   
    \author[a]{and Sebasti\'an C\'espedes}  
	\affiliation[a]{Theoretical Physics, Blackett Laboratory, Imperial College, London, SW7 2AZ, U.K }
    \emailAdd{m.carrillo-gonzalez@imperial.ac.uk}
    \emailAdd{s.cespedes-castillo@imperial.ac.uk}
\abstract{Effective field theories (EFTs) parametrize our ignorance of the underlying UV theory through their Wilson coefficients. However, not all values of these coefficients are consistent with fundamental physical principles. In this paper, we explore the consequences of imposing causal propagation on the comoving curvature perturbation in the EFT of inflation, particularly its impact on the primordial power spectrum and the effective sound speed  $c_s^\text{eff}$. We investigate scenarios where $c_s^\text{eff}$ undergoes a transition, remaining consistent with CMB constraints at early times but later experiencing a drastic change, becoming highly subluminal. Such scenarios allow the primordial power spectrum to grow at small scales, potentially leading to the formation of primordial black holes or the generation of scalar-induced gravitational waves. We find the generic feature that in a causal theory, luminal sound speeds imply a free theory, effectively constraining the dynamics. Additionally, we obtain that when considering natural values for the Wilson coefficients, maintaining the validity of the EFT and the weakly coupled regime, and enforcing causal propagation of the EFT modes, the power spectrum cannot increase drastically. This imposes significant constraints on the parameter space of models aiming to produce such features.}

\maketitle

\pagebreak
    
%%%%%%%%%%%%%%%%%%%%%%
\section{Introduction}
While Effective Field Theories (EFTs) are a powerful tool to characterize physics at low energies without the knowledge of a UV completion, it is well known that certain values of their Wilson coefficients can lead to unphysical properties such as breaking unitarity or causality~\cite{deRham:2022hpx,Kruczenski:2022lot}. When Lorentz invariance is not broken, there are well-developed techniques that allow us to put bounds on these coefficients by analysing scattering amplitudes and the analyticity requirements that physical properties impose on them~\cite{PhysRev.182.1400,Pham:1985cr, PhysRevD.31.3027,Pennington:1994kc,Ananthanarayan:1994hf,Comellas:1995hq,Manohar:2008tc,Adams:2006sv,deRham:2017avq,deRham:2017zjm,Bellazzini:2015cra,Bern:2021ppb,Bellazzini:2016xrt,Vecchi:2007na,Nicolis:2009qm,deRham:2022hpx}. On the other hand, if Lorentz invariance is broken, or if we want to probe interaction beyond 4-points, many of these techniques are no longer applicable. In the past few years, there has been a large focus on understanding how to apply similar ideas to settings without full Lorentz invariance, including cosmology \cite{Baumann:2022jpr}. These include explorations on the implications of unitarity~\cite{Grall:2020tqc, Melville:2021lst, Grall:2021xxm,Cespedes:2020xqq,Goodhew:2020hob,Goodhew:2021oqg,Jazayeri:2021fvk,Stefanyszyn:2024msm,Albayrak:2023hie} and causality~\cite{Benincasa:2020uph,Melville:2024zjq,Serra:2022pzl,Bittermann:2022hhy,Dubovsky:2007ac,Baumgart:2020oby,AguiSalcedo:2023nds}, the analyticity of correlation functions, wavefunctions, and scattering amplitudes \cite{AguiSalcedo:2023nds,Salcedo:2022aal,Baumann:2021fxj}, a definition of an S-matrix in cosmology \cite{Melville:2024ove,Melville:2023kgd,Albrychiewicz:2020ruh,Dvali:2017eba,Bousso:2004tv,Spradlin:2001nb,Cheung:2022pdk}, analysis of explicit examples with known and controlled UV completions \cite{Creminelli:2022onn,Creminelli:2023kze,Serra:2024tmz,Hui:2023pxc}, classical statistical inequalities \cite{PhysRevD.77.023505,Green:2023ids}, and many others.

Here, we will consider the implications of causality in terms of the support of the retarded Green's function \cite{deRham:2019ctd, deRham:2020zyh,Chen:2023rar,deRham:2021bll,Chen:2021bvg,Serra:2022pzl,Serra:2023nrn,CarrilloGonzalez:2022fwg,CarrilloGonzalez:2023cbf,CarrilloGonzalez:2023emp}. In flat spacetimes, with gravity turned off, the support of the Green's function can be easily defined in terms of the time delay, an observable that is well-defined in terms of the S-matrix: $\Delta T = -i \langle \text{in} | \hat{S}^\dagger \frac{\partial}{\partial \omega} \hat{S} | \text{in} \rangle \,$. Thus, we require that the Green's function does not have a resolvable support outside of the lightcone, that is, that the time delay satisfies $\Delta T \gtrsim - 1/\omega$, where $\omega$ is the frequency of the scattered wave (or the energy of the scattered particle, depending on the regime being probed). This inequality arises since a time advance smaller than $1/\omega$ cannot be measured due to the uncertainty principle. Note that this is not a strict inequality since the order one number on the right-hand side has only been formally probed in the context of quantum mechanics \cite{Wigner:1955zz}. We should also highlight that for the time delay to correspond to the support of the Green's function, we need to work with large enough frequencies\footnote{See Appendix A in \cite{CarrilloGonzalez:2022fwg} for a field theory example and \cite{Eisenbud, Wigner:1955zz} for a quantum mechanics proof of how the effective potential is probed at low-frequencies instead of the support of the Green's function.}. For example, when considering the scattering off a non-trivial background (which can also be thought of as the scattering off a particle or particles that generate that background) the scale $1/\omega$ should be smaller than the scale of variation of the background. This means that we should work in the semi-classical / WKB regime. When considering bounds on effective field theories, the frequencies will also be bounded from above to stay within the regime of validity of the EFT. It has been shown that causality bounds are consistent with other constraints such as positivity bounds in many scenarios \cite{CarrilloGonzalez:2022fwg,CarrilloGonzalez:2023cbf,deRham:2020zyh,Chen:2023rar,deRham:2021bll,Chen:2021bvg,Serra:2022pzl,Serra:2023nrn}.

If there is no conserved energy, a time delay can no longer be defined. Nevertheless, we can still probe the support of the retarded Green's function as follows. Consider an incoming wave packet that probes a non-trivial localized background configuration, this allows us to have boundary conditions akin to a scattering problem. Thus, within the WKB regime, the outgoing wave packet suffers a phase shift, $\delta$ with respect to a wave packet that does not interact with the non-trivial localized background configuration, that is, a high-energy, minimally-coupled mode (photon). This phase shift encodes the support of the retarded Green's function. If we had full Lorentz symmetry, the phase shift could be obtained as an eigenvalue of the S-matrix and its derivative with respect to $\omega$ will give the time delay. Given the symmetries of the problem, instead of a time delay, we should consider a spatial shift, $\Delta x = -\partial_k \delta$, where $k$ is the conserved spatial momentum, as done in \cite{deRham:2020zyh,CarrilloGonzalez:2023emp}. The causality bound is then given by
\begin{equation}
    \Delta r<1/k \ . \label{eq:causalbound}
\end{equation}
Theories satisfying this bound have a retarded Green's function that has no resolvable support outside the lightcone seen by high-energy modes.
In other words, we assume a spacetime whose chronology is determined by a high-energy mode, which determines the lightcone of comparison for the phase shift. If instead, the chronology of the spacetime is determined by an EFT mode such as in \cite{Bruneton:2006gf,Bruneton:2007si,Babichev:2007dw} then the reference lightcone would be different. The bound in Eq.~\eqref{eq:causalbound} also implies that modes have causal propagation in the sense that incoming wave packets always arrive at the scatterer (non-trivial background configuration) before the outgoing wave packets leave the scatterer \cite{Mizera:2023tfe,CarrilloGonzalez:2023emp}. Note that here we will not refer to other notions of violations of causality such as closed timelike curves\footnote{It is common to consider that a resolvable time advance can lead to closed timelike curves, but this has been previously contested in \cite{Burrage:2011cr} and recently \cite{Kaplan:2024qtf} which show that on a closed timelike curve the theory can become strongly coupled and the EFT breaks down. This is just an analogous version of the Hawking chronology protection \cite{Hawking:1991nk,Kim:1991mc} for EFTs.}.

In the Effective field theory of inflation~\cite{Creminelli:2006xe,Cheung:2007st} the Wilson coefficients of all operators can have large time variations, provided that the time scale is sufficiently slow that no new degrees of freedom are excited. At Cosmic Microwave Background (CMB) scales, time-dependent changes in the background are tightly constrained by observations \cite{Planck:2018jri}. However, at smaller scales, these constraints are not necessarily as stringent, allowing for larger variations in the parameters of the EFT of inflation. Such variations are phenomenologically interesting, as large peaks in the power spectrum can lead to the formation of primordial black holes or the generation of scalar-induced gravitational waves.

In the most commonly discussed scenarios where the power spectrum is enhanced, there is a brief phase in which the inflationary potential becomes nearly flat, reducing the value of  $\epsilon=\tfrac{\vert\dot H\vert}{H^2} $ and, crucially, driving  $\eta=\tfrac{d\epsilon}{dN}$  to be  $\mathcal{O}(1)$\footnote{See~\cite{Ozsoy:2023ryl} for a recent review.}. These scenarios typically require fine-tuning of the inflationary potential, necessitating a case-by-case study of higher-order corrections.

In this paper, we focus on scenarios in which the growth of the power spectrum arises from changes in the EFT parameters, where we modify the behaviour of the perturbations rather than modifying the potential \cite{Ballesteros:2018wlw,Ballesteros:2021fsp}. This approach allows for better control over the magnitude of higher-order interactions, providing a more systematic framework for studying these effects. Imposing that there are no resolvable positive spatial shifts allows us to constrain the space of parameters of the EFT. In particular, we find that when the speed of sound of the perturbations is luminal the theory is necessarily free, with all the Wilson coefficients set to zero, as was previously conjectured in \cite{Baumann:2015nta,Baumann:2019ghk}. In scenarios where the speed of sound decreases, we apply causality bounds that constrain how much the power spectrum can grow. This, together with the requirement of the EFT to be valid significantly constrains the space of parameters. 

The outline of this paper is as follows. We first review the EFT of inflation and the specific case under consideration, emphasizing the conditions under which this EFT remains valid. Next, we discuss the application of causality arguments to cosmology. In Section \ref{sec:caus_bounds}, we derive the causality bounds imposed on the EFT. In Section \ref{sec:2field}, we examine a specific two-field model to illustrate how when starting from a known physical UV completion, the EFT itself prevents causality violations. In Section \ref{sec:Conclusions}, we summarize our results and provide an outlook. Finally, we present a detailed description of our method to obtain causality bounds and further discussions on the strongly coupled regime in the Appendices.

%%%%%%%%%%%%%%%%%%%%%%
\section{Growth of power spectrum in EFT of inflation} \label{sec:PBH}

Let us consider the effective field theory (EFT) of a single scalar degree of freedom coupled to an FLRW metric, with broken time diffeomorphisms in unitary gauge~\cite{Creminelli:2006xe,Cheung:2007st}. Up to second order in perturbations, the action includes the following set of operators:
\begin{equation}
\begin{split}
S = \int d^4 x \sqrt{-g}  \bigg[
&	 \frac{\MP^2}{2} R - \MP^2(3H^2+\dot H) + \MP^2\dot H  g^{00} 
\\
&	+ \frac{1}{2}M_2^4 (\delta g^{00})^2
- \frac{1}{2}\hat{M}^3_1 \delta g^{00}\delta K   	 
%- \frac{1}{2}\bar{M}^2_1\left( \delta K_{\mu\nu}\delta K^{\mu\nu} -  \delta K^2\right)
  - \frac{1}{2}\bar{M}^2_2(\delta K)^2 
%  - \frac{1}{2}\bar{M}^2_3\delta g^{00} {}^{(3)} \! R
+ \ldots
\bigg] \, ,
\end{split}
\label{eftofi}
\end{equation}
where $K={K^\mu}_\mu$ denotes the trace of the extrinsic curvature $K_{\mu\nu}$ associated with the equal-time hypersurfaces of the spacetime foliation defined by the unitary gauge choice  \cite{Cheung:2007st}. The first line of the action contains only background terms, while the second line contains the term quadratic in the fluctuations. The coefficients $M_2^4$, $\hat{M}_1^3$ and $\bar{M}_2$ are Wilson coefficients that can be time-dependent due to the break of time diffeomorphism invariance. We will focus below on the action \eqref{eftofi} where the effective couplings satisfy the following hierarchy,\footnote{Notice that $H$ is the only relevant scale for the size of (massless) metric fluctuations during inflation. Therefore, since $\delta K$ has dimensions of mass one, its expected order of magnitude is $\delta K\sim H$ on purely dimensional grounds.}
\begin{equation}
M_2^4 \sim  \hat{M}^3_1H \gg \bar{M}^2_2H^2 \, ,
\label{eq:WBG}
\end{equation}
This choice might seem unusual, as one would typically expect the second operator, $(\delta g^{00} \delta K)$, to be suppressed compared to $(\delta g^{00})^2$. However, in theories with weakly broken Galileon  (WBG) symmetry, such a hierarchy can be achieved~\cite{Pirtskhalava:2015ebk, Goon:2016ihr,Santoni:2018rrx,Ballesteros:2021fsp}. Furthermore, this symmetry ensures that quantum corrections to these two operators are suppressed.

On the other hand, the operator $(\delta K)^2$ is not protected by any symmetry and is therefore expected to be subdominant compared to the other two. Nevertheless, we will still include it in our analysis, as it plays an important role by introducing quartic terms into the dispersion relation, which we aim to explore. Even though other operators appear at the same order in derivatives s $(\delta K)^2$, we will choose to ignore them as their effect will not change our conclusions. In this sense, the action of \eqref{eftofi} can be thought of as illustrative of this class of operators. Requiring that we remain within the regime of validity of an EFT with WBG implies that the values of the coefficients are restricted to ~\cite{Pirtskhalava:2015ebk, Goon:2016ihr,Santoni:2018rrx}, 
\begin{align}
    \epsilon M_{\mathrm{Pl}}^2 H^2\leq M_2^4,\ \hat{M}_1^3\leq M_{\mathrm{Pl}}^2 H^,\qquad \epsilon^{2/3}M_{\mathrm{Pl}}^{4/3}H^{2/3}\leq \bar{M}_2^2\leq M_{\mathrm{Pl}}^{4/3}H^{2/3} \ ,
\end{align}
where $\epsilon\equiv -\dot H/H^2\ll 1$ is the usual slow roll parameter. A further simplification can be done by working in the  regime where the scalar degree of freedom is decoupled from the gravitational fluctuations \cite{Cheung:2007st}, which in this case applies if ~\cite{Pirtskhalava:2015ebk},
\begin{align}
    M_2^4,\hat{M}_1^3 H\ll M_{\mathrm{Pl}}^2H^2,\qquad \bar{M}_2^2\ll M_{\mathrm{Pl}}^{4/3}H^{2/3} \ ,
    \label{decoupling}
\end{align}

To study the dynamics of the scalar  perturbations it is convenient to use the $\zeta$-gauge, defined by \cite{Maldacena:2002vr}:
\begin{equation}
\delta g_{ij} = a^2 e^{2\zeta} \delta_{ij}\,.
\end{equation} 
After integrating out the non-dynamical components of the metric, $\delta g^{00}$ and $\delta g^{0i}$, from \eqref{eftofi}, and in the decoupling limit one finds the following quadratic  action for $\zeta$,
\begin{equation}
S^{(2)}_\zeta = \int d^4 x \, a^3 H^{-2} \left(2 M_2^4 - \MP^2\dot H \right)  \left[  \dot \zeta^2 - c_s^2 \frac{(\partial_i\zeta)^2}{a^2}  - \alpha \frac{(\partial_i^2\zeta)^2}{H^2a^4}\right] \, ,
\label{ghostL2}
\end{equation}
where 
\begin{equation}
%c_s^2 \equiv \frac{-2\left(\MP^2 + 3\bar{M}_2^2 \right)\dot H  +  \hat{M}_1^3 H  + \partial_t (\hat{M}_1^3) }{ 2( 2 M_2^4 - \MP^2\dot H )} \, ,
c_s^2 = \frac{-2\MP^2 \dot H  +  \hat{M}_1^3 H  + \partial_t (\hat{M}_1^3) }{ 2( 2 M_2^4 - \MP^2\dot H )} \, ,
\qquad
\alpha = \frac{\bar{M}_2^2H^2}{2(2 M_2^4 + \MP^2\vert\dot H\vert )} \, .
\label{cs}
\end{equation}
In general both $c_s^2$ and $\alpha$ receive additional contributions from the operators we have ignored, however, these are subleading in the decoupling limit due to \eqref{decoupling}. Additionally, $\alpha$ is expected to always receive positive contributions when all the Wilson coefficients are real. It is possible to make an additional simplification by assuming that the inflationary dynamics is predominantly driven by the potential, rather than the kinetic term. This choice makes it more natural to take the decoupling limit since, when the theory is kinetically driven, the hierarchies in \eqref{decoupling} do not necessarily hold. The choice of inflation driven by the potential  corresponds to the following conditions~\cite{Pirtskhalava:2015nla, Pirtskhalava:2015zwa}:
\begin{equation}
M_2^4 \sim  \hat{M}^3_1H  \sim \varepsilon \MP^2H^2  \, ,
\qquad\qquad
\bar{M}_2^2  \sim \varepsilon^{2/3} \MP^{4/3}H^{2/3}  \, ,
\label{M2Mb2h}
\end{equation}
Notice that this implies that $\alpha$ cannot be arbitrarily small without fine-tuning. This will be important in what follows as we will vary the parameters $\alpha$ and $c_s$ to probe different regimes of the theory. \\

\noindent \textbf{Equations of motion} 
It will be convenient to change coordinates to $v=z\zeta$ with,
\begin{align}
    z^2\equiv \frac{2a^2}{H^2}\left(2M_2^4-M_{\mathrm{pl}}^2\dot H\right) \ ,
\end{align}
with this change of coordinates, the action becomes,
\begin{align}
    S^2_v=\frac{1}{2M^2}\int d^3xd\tau\left[(v')^2-c_s^2(\partial_i v)^2-\frac{\alpha}{H^2a^2}(\partial_i^2v)^2+\frac{z''}{z}v\right]  \ .\label{eq:Swithz}
\end{align}
We also have that,
\begin{align}
    \frac{z''}{z}=a^2H^2\left(\left(1+\frac{1}{2}\left(\eta-2\gamma\right)\right)\left(2-\epsilon+\frac{1}{2}(\eta-2\gamma)\right)+\frac{1}{2}\frac{d}{dN}(\eta-2\gamma)\right) \ , \label{eq:zder}
\end{align}
where the corrections of order $\epsilon$ are derivatives of the background. These can be larger than one but its effect will be important after horizon crossing as seen from the time dependence of the term. After performing a Fourier transform and ignoring slow roll corrections, the equation of motion reads,
\begin{align}
    v_k''+\left(c_s^2k^2+\alpha k^4\tau^2-\frac{2}{\tau^2}\right)v_k=0 \ .
    \label{MSequation}
\end{align}
It will be useful to find solutions for both cases when either $\alpha=0$ or  $c_s=0$. For the former, the positive frequency solution in the usual Bunch-Davies vacuum is given by,
\begin{align}
    v_k=\frac{1}{\sqrt{2c_s k}}\frac{1+i c_sk \tau}{c_s k\tau}e^{-ik \tau} \ .
\end{align}
The amplitude of the power spectrum is given by,
\begin{align}
    \Delta^2_\zeta\equiv\frac{k^3}{2\pi^3}P_\zeta=\lim_{k\vert\tau\vert\to 0}\vert v_k/z\vert^2=\frac{H^2}{4\pi^2}\frac{\vert\dot H\vert}{2M_{\mathrm{Pl}}^2\vert \dot{H}\vert+\hat{M}_1^3 H}\frac{1}{\epsilon c_s} \ .
    \label{PS_cs}
\end{align}
The case $c_s=0$ corresponds to ghost inflation~\cite{ArkaniHamed:2003uy}\footnote{A general analytic solution of the equation can be obtained. In this case, the positive frequency solution  for the mode function is~\cite{Gorji:2021isn},
\begin{align}
v_k=\left(\alpha\right)^{-1/4}\frac{e^{-\frac{\pi c_s^2}{8\alpha^{1/2}}}}{\sqrt{-2k^2\tau}} W\left(\frac{ic_s^2}{4\alpha^{1/2}},\nu/2,-i\alpha^{1/2}k^2\tau^2 \right) \ ,
\end{align}
where $W$ is the Whittaker function. This function reproduces the usual solution for $\alpha=0$.}. The normalised positive frequency solution is given by,
\begin{align}
    v_k=\sqrt{\frac{\pi}{8}} \sqrt{-\tau}H_{3/4}^{(1)}\left(\frac{\sqrt{\alpha}}{2}k^2\tau^2\right).
    \label{SolGhostinfl}
\end{align}
Notice that in the superhorizon limit, the solution \eqref{SolGhostinfl} does not match the usual flat space solution. This happens since at early times, the dispersion relation is non-relativistic $\omega^2\sim\alpha k^4/a^4$. Nevertheless, the vacuum that defines the solution is analogous to the usual Bunch-Davis in the sense that it is a positive-frequency solution without particles. Moreover, the amplitude of the power spectrum is also scale-invariant on superhorizon scales:
\begin{align}
    \Delta^2_\zeta=\lim_{k\vert\tau\vert\to 0}\vert v_k/z\vert=\frac{\Gamma(3/4)^2}{4\pi^3}\frac{H^4}{\vert\dot H\vert M_{\mathrm{Pl}}^2+2M_2^4}\frac{1}{\alpha^{3/4}} \ .
    \label{PS_alpha}
\end{align}
\\ 

\noindent \textbf{Weakly coupled regime} 
As with any EFT, the one we consider should be understood as the low-energy limit of a more fundamental theory that remains valid up to arbitrarily high energies. Since the underlying theory is not directly accessible and our calculations are perturbative, we assume that while the EFT may remain valid up to a cutoff scale $\Lambda_{\mathrm{UV}}$, our results should only be trusted up to the strong coupling scale $\Lambda_\star$, beyond which perturbation theory breaks down.  In general, this scale  $\Lambda_\star$  is lower than the cutoff  $\Lambda_{\mathrm{UV}}$\footnote{See Appendix \ref{app:sc} for details.}.

Given an EFT consisting of a series of non-linear derivative interactions, the strong coupling scale corresponds to the lowest energy at which one of these operators becomes large enough that the weak coupling expansion is no longer valid. In practical terms, this restricts the size of the Wilson coefficients, which, in the case of interest, translates into constraints on the parameters  $c_s$  and  $\alpha$. For instance, in the action \eqref{eftofi}, the operator  $(\delta g^{00})^2$  contains the cubic terms  $(\nabla\zeta)^2 \dot\zeta$  and  $\dot\zeta^3$ . As the coefficient  $M_2^4$  increases, both the speed of sound  $c_s$  and the parameter  $\alpha$ decrease. Which term dominates the dispersion relation plays an important role in determining the first operator that becomes strongly coupled.  In the case when $c_s$ dominates over $\alpha$  the relevant energy scale is given by,
\begin{align}
    \Lambda_2^4\simeq 2\pi^2 c_s^5\vert\dot H\vert (1+\alpha_2)^3\alpha_2^{-2}, \qquad \mathrm{with}\  \alpha_2=\frac{\hat{M}_1^3H}{2M_{\mathrm{Pl}}^2\vert\dot H\vert} \ ,
\end{align}
Other operators will have larger energy scales than this, so we find that in this case, the strong coupling scale is   $\Lambda_\star=\Lambda_2$. Now perturbation theory remains valid as long as the strong coupling scale is below the scale of inflation $H$. Recalling the assumption of potential-dominated dynamics in Eq.~\eqref{M2Mb2h}, we note that \( \alpha_2 \) is typically of order one. Under this condition, the energy scale reduces to that found in \cite{Cheung:2007st}, namely  
\be
\Lambda_\star=\Lambda_2^4 \simeq 2\pi^2 H^4\Delta_\zeta^{-2} c_s^4 \ ,
\ee
which imposes the following constraint on the speed of sound:  
\be
2\pi^2 c_s^4 \gg \Delta_\zeta^2 \ ,
\ee
where we have used the expression for the power spectrum given in Eq.~\eqref{PS_cs}. In the case when the $\alpha$ term dominates in the dispersion relation, the scaling of the energy scales changes and the first operator to become strongly coupled is given by,
is  
\[
\frac{\hat{M}_1^3}{H^3} \frac{1}{a^4} (\partial_i^2\zeta)(\partial_j\zeta)^2 \ .
\]
The energy scale associated with this operator can also be estimated through dimensional analysis using the modified dispersion relation  $\omega^2 = \frac{\alpha k^4}{H^2}$, which leads to  
\[
\Lambda_3 \sim \frac{H}{\sqrt{\alpha}}.
\]
This energy scale gives the strong coupling scale $\Lambda_\star=\Lambda_3$, since due to the different scaling all other energy scales become parametrically larger when $\alpha\ll 1$, for example $\Lambda_2\sim\Lambda_3\alpha^{-2}\gg\Lambda_3$.

Finally, it is important to note that these constraints depend on the weakly broken Galileon (WBG) symmetry and the assumption in Eq.~\eqref{M2Mb2h}. Different symmetry choices could lead to other operators becoming strongly coupled. Nevertheless, the model we have chosen remains robust under loop corrections, which significantly simplifies the analysis by allowing us to work in the weakly coupled regime. \\

\noindent \textbf{Regime of validity of the EFT} 
Let us note that higher-order derivative corrections in principle modify Eq.\eqref{MSequation} by introducing additional terms with higher powers of the physical momentum  $k/a$. However, these corrections can be safely assumed to be suppressed, provided that the constraints imposed by the WBG symmetry in Eq.\eqref{eq:WBG} are satisfied and that
\begin{align}
\sqrt{\alpha} \frac{k}{aH} \ll 1.
\label{alphacond}
\end{align}
Further constraints exist on how much Wilson coefficients can vary with time, ensuring that the background remains adiabatic. Specifically, the EFT should resolve time variations that are longer than the minimal timescale dictated by the corresponding strong coupling scale  $\Lambda_\star^{-1}$. Estimating the time variation of a function  $f$  over an  e-fold as $H^{-1} \frac{d f}{dt} \sim \left(\frac{\Gamma}{H}\right) f$ means that we should require 
\be
  \frac{\Gamma}{H}\ll 1 \ .
\ee
In the present case, we will estimate the time variations as follows
\begin{equation}
    \frac{\Gamma}{H}\equiv\frac{\Delta \partial_N c_s^2}{\Delta c_s^2}\ ,  \quad \text{where} \quad \Delta f = \left|\lim_{N\rightarrow - \infty}f(N)-\min(f) \right|\ ,\label{eq:GoverH}
\end{equation}
where $\min(f) $ is the minimum of the function $f$. Note that this is well defined in the limit where $c_s$ vanishes and, in the parametrizations of the transition that we will consider below, it also captures the time evolution of $\alpha$. If we further assume that the strong coupling scale is still given by $\Lambda_\star$, we also require that $\Gamma\ll \Lambda_\star$. 

\subsection{Large Power spectrum and PBHs}
We are interested in mechanisms that allow the power spectrum to grow significantly over a few e-folds while remaining within the regime of validity of the effective field theory. This is particularly intriguing because a transient peak in the power spectrum can lead to the formation of primordial black holes (PBHs) or induce the production of gravitational waves. In such scenarios, for the signal to be detectable, the power spectrum must increase by at least six orders of magnitude from its CMB value~\cite{Ozsoy:2023ryl}.  

Several approaches have been proposed to achieve this enhancement while ensuring the EFT remains valid~\cite{Ballesteros:2018wlw, Ballesteros:2021fsp}. One natural method is by decreasing the speed of sound \( c_s \), using the fact that when $\alpha=0$ and $c_s$ are constant, the power spectrum is given by 
\begin{equation}
\Delta_\zeta^2 = \frac{1}{8\pi^2}\frac{H^4}{M_{\mathrm{Pl}}^4 \epsilon c_s}.
\end{equation}

However, as we have discussed, this enhancement is constrained by strong coupling scale $\Lambda_\star$, which requires \( c_s \leq \Delta_\zeta^{1/2} \). This limits its maximum value to around \( 10^{-5} \), preventing sufficient growth ~\cite{Ballesteros:2018wlw}. As we have seen, this issue can be circumvented if additional operators contribute at quadratic order in the equations of motion. If an alternative quadratic operator controls the strong coupling scale $\Lambda_\star$, this scale is no longer dictated by \( c_s \) and the power spectrum can experience greater growth. This is, of course, the case when $\alpha$ dominates.  In such scenarios, we expect an enhancement of the power spectrum by a factor of  \( \alpha^{-3/4} \), consistent with the scaling derived in \eqref{PS_alpha} for a constant $\alpha$. Even though staying in the weak coupling regime does not impose a strict lower bound on  $\alpha$, assuming weakly broken Galileon symmetry leads to the relation
\begin{equation}
\alpha \sim \left(\frac{H}{M_{\mathrm{Pl}}}\right)^{2/3} \epsilon^{-1/3},
\label{natural_alpha}
\end{equation}
which shows that achieving very small values of  $\alpha$  would require fine-tuning higher-order derivative operators—e.g., $(\delta g_0)^3$ and similar terms—that would otherwise generate, through loop corrections, an  $\alpha$  coefficient of the magnitude given in \eqref{natural_alpha}. Additionally, since the power spectrum must remain consistent with CMB constraints at large scales, we require that in the EFT we are considering, the theory is dominated by $c_s$ at CMB scales, and that for shorter scales $\alpha$ dominates. This implies that at the time of the transition between the two phases, $t_*$, the strong coupling scales $\Lambda_\star$ changes from $\Lambda_2$ to $\Lambda_3$, which leads to the condition,
\begin{align}
\frac{1}{\alpha(t_*)}\lesssim     c_s^2(t_*)\Delta_\zeta^{-1}(t_*).   
\end{align}
which is in general a stronger constraint than assuming weak coupling ($\alpha \ll 1$).  On top of this, we need to consider that sudden variations in the power spectrum must also satisfy adiabaticity to ensure the validity of the EFT. Taken together, these constraints suggest that generating a peak large enough to form PBHs within this EFT framework is challenging. In \cite{Ballesteros:2018wlw}, the adiabaticity condition was relaxed by assuming an instantaneous transition. In the case when $\alpha$ dominates after a transition where $c_s$ dominates the power spectrum is given by,
\begin{equation}
\Delta_\zeta^2 \simeq \frac{4\pi}{\Gamma(1/4)^2}\frac{c_s^3}{\alpha^{3/4}}\Delta_\zeta^2(k\to 0)
\label{PS_ST}
\end{equation}
where $c_s$ and $\Delta_\zeta^2(k\to 0)$ are the speed of sound and the power spectrum before the transition. In this idealised scenario, the maximum growth is still given by the value of the power spectrum with $\alpha$ constant. Even though, in a realistic scenario, this dependence will be more complicated, the expression for the power spectrum in \eqref{PS_ST} gives an estimate of how large it can grow.

%%%%%%%%%%%%%%%%%%%%%%
\section{Causality and spatial shifts}
We proceed to analyse the requirements of causality on the propagating modes of an EFT in a curved background; these ideas have been explored in \cite{Dubovsky:2007ac,deRham:2019ctd, deRham:2020zyh,Baumgart:2020oby,Serra:2022pzl,Bittermann:2022hhy,CarrilloGonzalez:2023emp,Melville:2024zjq,Hui:2025aja}. Let us start by considering a wave packet that describes the propagation of an EFT mode on a non-trivial background. As this mode propagates and interacts with / scatters off the background, it will obtain a phase shift $\delta$ with respect to a high-energy mode in the same background. Thus, the outgoing wave packet has the shape
\be
\phi^\text{out}\sim \int d^3{\bm {k}} \ A_k \ e^{i {\bm k}\cdot{\bm x}}e^{-i\omega t}e^{i 2 \delta} \ .
\ee
We will assume that the high-energy mode defines the chronology of the spacetime. For example, we can think of it as a minimally coupled photon in the inflationary background. Additionally, we will impose that in the far past and far future, the EFT interactions decay fast enough so that the mode propagates in the same manner as the high-energy mode, that is, its propagation only changes due to the spacetime curvature. This allows us to set up an analogy of a scattering problem to have a well-defined phase shift due to a localized scattering region. Another requirement for having a well-defined phase shift is that the length scales of the background ($1/H$) are much larger than those of the propagating mode ($k/a$). In other words, we will consider modes within the horizon, which can be described by the semi-classical (WKB) approximation. We can see that the centre of the wave packet arrives at a fixed time slice $t=t_f$ at a distance $x(t_f)-2\partial_k\delta$ where $x(t_f)$ is the distance travelled by the high-energy mode. Thus, the EFT mode suffers a physical spatial shift $\Delta_r$ given by
\begin{equation}
    a(t_f) \Delta r=-2a(t_f)\frac{\partial\delta}{\partial k} \ . \label{eq:Dr}
\end{equation}
From this point of view, a positive spatial shift implies that the wave packet left the scattering region before it arrived at it, which violates causality. Due to the uncertainty principle, spatial delays smaller than the physical wavelength $a(t_f)/k$ are not resolvable. Hence, the constraint to have causal propagation is
\begin{equation}
\frac{k}{a(t_f)}(a(t_f)\Delta r)\leq 1 \ .  \label{eq:causal}
\end{equation}
In the present analysis, we consider modes that satisfy the equation of motion in Eq.~\eqref{MSequation}. We can do a further change of variables from conformal time to e-folds $N$, defined as $\d N = H \d t = a H \d\tau$, to find
\begin{equation}
\frac{\d^2 v_k}{\d N^2 } + (1-\varepsilon) \frac{\d v_k}{\d N } 
+ \left[\left(c_s^2+\frac{\alpha  k^2}{a^2H^2}\right)  k^2-\frac{z''}{z}
\right]\frac{v_k }{a^2 H^2 } =0 \, .
 \label{eq:eomN}
\end{equation} 
where $z''/z$ is given in Eq.~\eqref{eq:zder}. Last, we perform a field redefinition
\begin{equation}
    \Psi_k=e^{\frac{1}{2}(1-\epsilon)N}v_k \ ,
\end{equation}
to get rid of the friction term so that the equation of motion now reads
\begin{align}
\frac{\d^2 \Psi_k}{\d N^2 }&+  W_k(N)\Psi_k =0 \,  \label{eq:eomNnofric}  , \\
W_k(N)&=\frac{k^2}{a^2 H^2 }\left(c_s^2+\alpha \frac{ k^2}{a^2H^2}-\frac{1}{4}\frac{ a^2H^2}{ k^2}\left(3+\gamma-\epsilon+\eta\right)^2 \label{eq:W}
\right) \ .
\end{align} 
We can see that as long as we consider modes inside the horizon, that is, satisfying
\begin{equation}
\frac{k}{ aH}\gg 1 \ , \label{eq:wkb}
\end{equation}
we can solve this equation using the WKB approximation and obtain the phase shift experienced by the EFT modes. Considering a Bunch-Davies vacuum in the far past, we find 
\begin{equation}
\Psi_k(N)\simeq A_k(N) e^{-i \int_{N_\text{ini}}^{N}\delta_{\mathrm{WKB}}(\tilde{N}) \d \tilde{N}} \ , \label{eq:WKBmodes}
\end{equation}
where at leading order in the WKB approximation the phase is $\delta_{\mathrm{WKB}}=\sqrt{W_{k}}$. One can include higher-order WKB corrections straightforwardly. The corrections will be suppressed by powers of $aH/k$, but can arise at the same order as the effective potential terms in Eq.~\eqref{eq:W}. The precise expressions, including higher-order $(aH/k)^2$ contributions are
\begin{equation}
\begin{aligned}
	&\delta_{\mathrm{WKB}}^{(0)}=\sqrt{W_{k}},\\
	&\delta_{\mathrm{WKB}}^{(2)}=- \frac{1}{8 \sqrt{W_{k}}}\left(\frac{W_{k}^{\prime \prime}}{W_{k}}-\frac{5}{4}\left(\frac{W_{k}^{\prime}}{W_{k}}\right)^2\right) \text {, }\\
	&\delta_{\mathrm{WKB}}^{(4)}=\frac{1}{32 W_{k}^{3 / 2}}\left[\frac{W_{k}^{(4)}}{W_{k}}-7 \frac{W_{k}^{\prime} W_{k}^{(3)}}{W_{k}^2}-\frac{19}{4}\left(\frac{W_{k}^{\prime \prime}}{W_{k}}\right)^2+\frac{221}{8} \frac{W_{k}^{\prime \prime} W_{k}^{\prime 2}}{W_{k}^3}-\frac{1105}{64}\left(\frac{W_{k}^{\prime}}{W_{k}}\right)^4\right] \ ,
\end{aligned}
\end{equation}
where primes denote derivatives with respect to e-folds. Defining the phase shift with respect to a high-energy mode in an FLRW universe gives
\begin{equation}
\delta(N)=\int_{N_{\text{ini}}}^{N}\left(\sum_{j \geq 0} \delta_{\mathrm{WKB}}^{(j)}(\tilde{N})-\sum_{j \geq 0} \delta_{\mathrm{WKB-FLRW}}^{(j)} (\tilde{N})\right)\mathrm{d} \tilde{N} \ , \label{eq:phaseshift}
\end{equation}
where 
\begin{equation}
\delta_{\mathrm{WKB-FLRW}}=\delta_{\mathrm{WKB}}\Big|_{c_s\rightarrow1, \alpha\rightarrow0} \ .
\end{equation}
Note that the $(j)$ term has a leading contribution of order $(k/(aH))^{1-j}$ so that higher orders in the WKB expansion are suppressed. 

At leading order, we can rewrite the causality requirement from Eq.~\eqref{eq:causal} as
\begin{equation}
k \left( \int_{N_\text{ini}}^{N_\text{final}}\left(\partial_k\omega(k,N) -\frac{1}{aH}\right)\d N \right)\leq 1 \ , \label{eq:cboundIntegrand}
\end{equation}
where $\omega$ is given by the dispersion relation for the modes satisfying Eq.~\eqref{eq:eomNnofric}
\begin{equation}
\omega^2(k,N)=c_s^2(N)\frac{ k^2}{a^2H^2}+\alpha(N) \frac{ k^4}{a^4 H^4} \equiv    ( c_s^\text{eff}(k,N))^2 \frac{ k^2}{a^2H^2}\ ,
    \label{eq:dispersion}
\end{equation}
where $ c_s^\text{eff}$ is the effective sound speed. We can further rewrite this bound by going back to conformal time and changing variables to $q=k/(aH)$,
\begin{equation}
k\int_{\tau_\text{ini}}^{\tau_\text{final}}\partial_q\omega(q) \d \tau \leq k (\tau_\text{final}-\tau_\text{ini}) \ .
\label{eq:causal2}
\end{equation}

Before obtaining explicit causality bounds, we will present another point of view on why the phase shift is related to the causality properties of an EFT in curved backgrounds. This has been previously pointed out in \cite{deRham:2019ctd,deRham:2020zyh}. Here, we will argue that one can think of the phase shift as encoding the support of the retarded Green's function. As a first approach to understanding this, one can compute the Green's function perturbatively in the EFT expansion, but this will not capture the physics of the semi-classical approximation, that is, the implicit resummation in the WKB exponential\footnote{Note that this resummation can be troublesome if it is not performed carefully. One should include all terms up to the desired order in the EFT in the phase shift so that the WKB approximation gives the correct result up to the chosen EFT order. These terms can arise from higher-order EFT expansions in the equations of motion and higher-order WKB corrections.}. For example, the leading order correction to the Green's function of the free theory, $G_{\mathrm{ret}}^0$, is 
\begin{equation}
	G_{\mathrm{ret}}^1\left(N, N^{\prime}\right)=G_{\mathrm{ret}}^0\left(N, N^{\prime}\right)-\int_{N^{\prime}}^N \mathrm{d} N^{\prime \prime} G_{\mathrm{ret}}^0\left(N, N^{\prime \prime}\right)\left(W_k\left( N^{\prime \prime}\right)-\left(\frac{k}{aH}\right)^2\right)  G_{\mathrm{ret}}^0\left(N^{\prime \prime}, N^{\prime}\right) \ .
\end{equation}
Since $G_{\mathrm{ret}}^0\left(t, t^{\prime}\right)$ has support within the FLRW lightcone, then $G_{\mathrm{ret}}^1$ will also have support in the same region. Adding higher-order EFT corrections will not change this conclusion. This result can change if we can resum the contributions up to a chosen EFT order due to their secular growth, which happens only when we have an observable spatial shift. In the setting described above, this is equivalent to computing the retarded Green's from its exact expression in momentum space,
\begin{equation}
	G_{\mathrm{ret}}\left(t, t^{\prime}\right)=\theta \left(t-t^{\prime}\right) i \left(\phi_k(t) \phi_k^*\left(t^{\prime}\right)-\phi_k\left(t^{\prime}\right) \phi_k^*(t)\right) \ , 
\end{equation}
but now with the positive frequency modes $\phi_k$ given by the WKB solution in Eq.~\eqref{eq:WKBmodes} which is the approach first proposed in \cite{Bunch:1979uk}. The exponential contribution arising from the WKB approximation involves the resummation of the corrections to the Green's function up to a given EFT order and reads
\begin{equation}
	\phi_k \sim \frac{e^{i k \tau}}{\sqrt{\omega(k,N)}} e^{-i k \left( \int_{-\infty}^{\tau}\left(c_s^\text{eff}-1\right)\d \tilde{\tau} \right)} \ ,
    \label{modefunctionWKB}
\end{equation}
at leading order. This implicit resummation is correct when the term in the second exponential is of order one or larger which, at leading order, is the case if there is a resolvable spatial shift. While it is not possible to directly compute the support of the Green's function in coordinate space without information of the theory in the UV to perform the inverse Fourier transform, it should be clear that the presence of the phase shift in the exponential will change the support of $G_\text{ret}$.

Note that similar ideas have been analysed recently in \cite{Hui:2025aja}. There, it was pointed out that since the retarded propagator  $\tilde{G}_{\mathrm{ret}}(t,\bm{x})$  is causal, it must be a distribution with compact support over a ball of radius $ R = |t| $. In the context of cosmology, this implies that its Fourier transform $G_{\mathrm{ret}}(\tau_1,\tau_2,k)$ is analytic in $k^2=\bm{k}\cdot\bm{k}$ and is bounded by
\begin{align}
    \vert G_{\mathrm{ret}}(\tau_1,\tau_2,k)\vert\leq C(D+\vert k\vert)^N e^{\vert
    \mathrm{Im}\ k \vert \  \vert \tau_2-\tau_1\vert} \ , \label{eq:Greenbound}
\end{align}
for some $C,D$ and $N$ positive constants. This bound is valid for wavenumbers within the regime of validity of the EFT. We can verify that \eqref{eq:Greenbound} is satisfied in our examples. Replacing  the mode function from  \eqref{modefunctionWKB} and keeping only the exponentials we find that,
\begin{align}
    G_{\mathrm{ret}}(\tau_1,\tau_2,k)\sim \mathrm{Im}\left(i \ e^{i (k (\tau_2-\tau_1))}e^{-ik (\int^{\tau_2}_{\tau_1}(c_s^{\mathrm{eff}}(k,\tilde{\tau})-1) d\tilde{\tau}}\right)\ .
\end{align}
If we take the norm of the retarded Green function we then find that,
\begin{align}
    \vert G_{\mathrm{ret}}(\tau_1,\tau_2,k)\vert\leq e^{\vert
    \mathrm{Im}\ k \vert \int^{\tau_2}_{\tau_1} (c_s^{\mathrm{eff}}(k,\tilde{
    \tau})-1) \d\tilde{\tau}} e^{ \vert
    \mathrm{Im}\ k \vert \  \vert \tau_2-\tau_1\vert}\ . \label{eq:greens_phaseshift}
\end{align}
As mentioned before, the implicit resummation in the exponential is valid only if the exponent is of order one which, at leading order, would lead to a resolvable spatial delay. If the exponent is smaller than one, the exponential should be series-expanded. In such cases, the spatial shift will not be resolvable and the retarded propagator will satisfy \eqref{eq:Greenbound}. Similarly, if the exponent is negative, which at leading order leads to a negative spatial shift, the retarded propagator will satisfy \eqref{eq:Greenbound}. This shows that our (leading order) causality requirement in Eq.~\eqref{eq:causal2} is equivalent to \eqref{eq:Greenbound}. When higher-order momentum terms dominate the sound speed, this argument requires more care since the momentum derivative of the phase shift's integrand is no longer proportional to itself. This allowed us to relate the exponent in Eq.~\eqref{eq:greens_phaseshift} to the spatial shift at leading order. In the present case, we can see that when $\alpha$ dominates the effective sound speed,  $\partial_k \delta^{(0)}_\text{WKB}\sim 2 c_s^\text{eff}+2 \alpha q^2 /c_s^\text{eff}\sim 2 c_s^\text{eff}+2 \sqrt{\alpha} q$. Since $\sqrt{\alpha} q\ll 1$ due to the validity of the EFT, then $\partial_k \delta^{(0)}_\text{WKB}\sim \mathcal{O}(1) \delta^{(0)}_\text{WKB}$ and the causality criteria in Eq.~\eqref{eq:causal} will again be equivalent to \eqref{eq:Greenbound}.

%%%%%%%%%%%%%%%%%%%%%%
\section{Causality bounds on the EFT of inflation}
\label{sec:caus_bounds}
\begin{figure}[!hbt]
    \centering
    \includegraphics[width=0.45\linewidth]{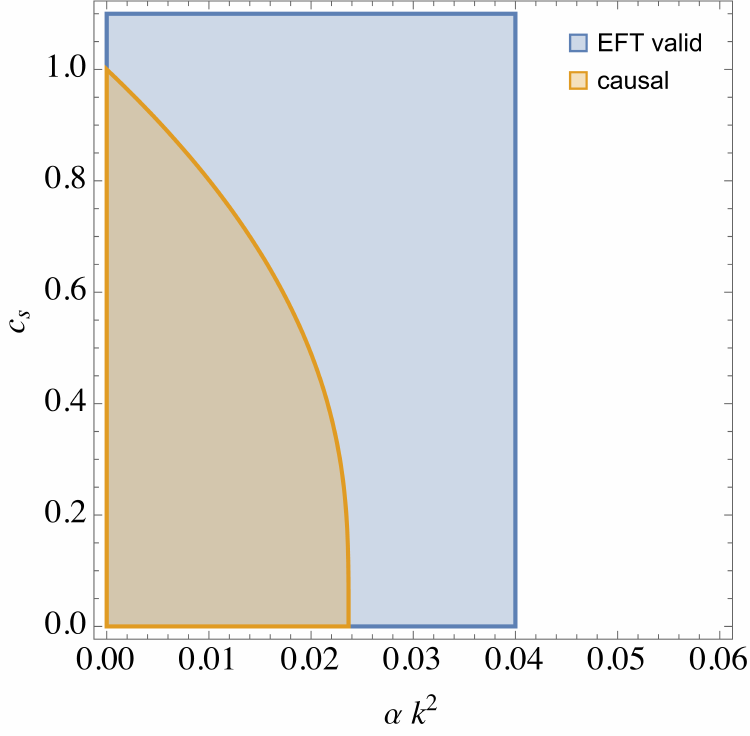}
    \caption{The yellow region shows the allowed parameter space by the causality bounds from \eqref{eq:cbound}. The blue region corresponds to the parameter space where the EFT is valid, $|\alpha k^2 \tau_{\mathrm{ini}}^2|<1$. We have set $\tau_{\mathrm{final}}=-1.5$ and $\tau_{\mathrm{ini}}=-5$.}
    \label{fig:CausalityGhostInfl}
\end{figure}

In this section, we will apply the causality bounds described above to constrain the effective sound speed of the comoving curvature perturbation, Eq.~\eqref{eq:dispersion}. We start analysing a simple case where $c_s$ and $\alpha$ are constant and the background is de Sitter. In this simplified scenario, the relation from  \eqref{eq:causal} becomes,
\begin{align}
    \frac{\tau_{\mathrm{final}}(c_s^2+\alpha k^2\tau_{\mathrm{final}}^2)^{1/2}-\tau_{\mathrm{ini}}(c_s^2+\alpha k^2\tau_{\mathrm{ini}}^2)^{1/2}}{\tau_{\mathrm{final}}-\tau_{\mathrm{ini}}}\leq1\ .
    \label{eq:cbound}
\end{align}
The validity of the EFT requires that we work at conformal times satisfying both the EFT constraint~$|\alpha k^2 \tau_{\mathrm{ini}}^2|\ll 1$ and the WKB approximation $k\tau\geq-1$. The bounds implied by Eq.~\eqref{eq:cbound} can be seen in Fig.~\ref{fig:CausalityGhostInfl}. When $c_s\simeq 1$ we can simplify Eq.~\eqref{eq:cbound} as
\begin{equation}
\frac{2 c_s^2 + k^2 (t_f^2 + t_f t_i + t_i^2) \alpha}{2 \sqrt{c_s^2}} < 1 \ .
\end{equation}
In particular, if $c_s=1$ the relation can only be satisfied for $\alpha=0$. This means that any higher-order correction has to vanish if the speed of sound is luminal. We will see that this feature holds in more general scenarios.

After looking at the simple example above, we will now look for bounds on the time (e-fold) dependence of $c_s$ and $\alpha$ which can be thought of as a parametrization of the time evolution of the Wilson coefficients of the EFT of inflation, see Eq.~\eqref{cs}, and hence of the cosmological background. We will consider the case where there is a transition period at early times that can lead to an enhancement of the power spectrum \cite{Ballesteros:2018wlw} as described in Section~\ref{sec:PBH}. In the following, we will analyse three parametrizations of this time dependence to understand the different features that influence the causal propagation, the validity of the EFT, and the growth of the power spectrum. In each case, we apply the bounds from Eq.~\eqref{eq:causal}, the specific method used to obtain the bounds is explained in detail in Appendix \ref{ap:method}. In all cases, we will assume that in the far past the effective sound speed is luminal and there is a Bunch-Davies vacuum. At some early time, the effective sound speed changes and can become superluminal for a finite period of time, before it becomes luminal again. We only consider modes for which the transition is over before leaving the horizon, this ensures that the WKB approximation holds and we have a well-defined phase shift. Additionally, the EFT should be valid for the whole duration of the transition. In other words, Eq.~\eqref{alphacond} and \eqref{eq:GoverH} should be satisfied till the end of the transition.

%%%%%%%%%%%%%%%%%%%%%
\subsection{Gaussian parametrization} \label{subsec:gauss}
\begin{figure}[!ht]
    \centering
    \includegraphics[width=0.42\linewidth]{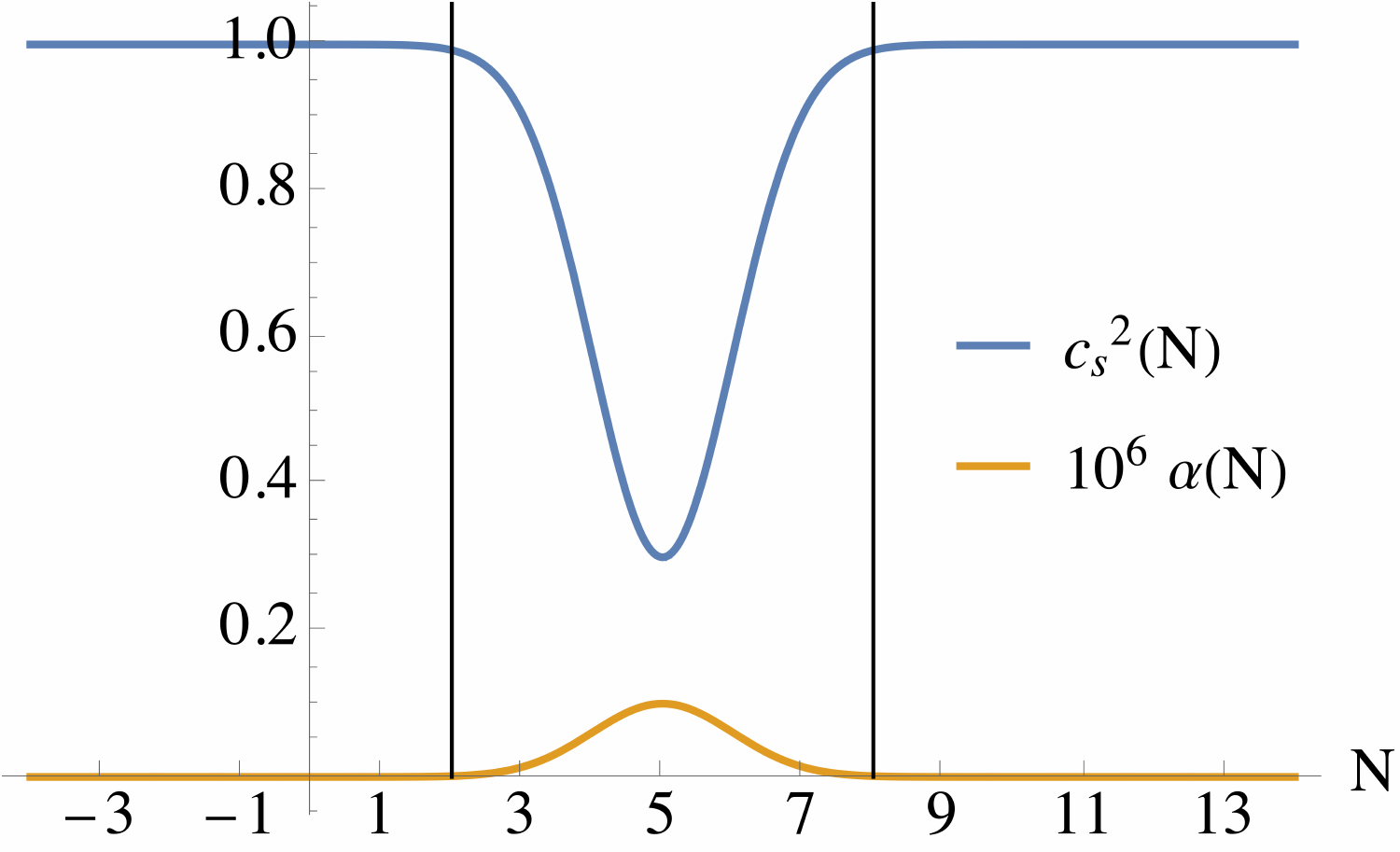}
    \includegraphics[width=0.56\linewidth]{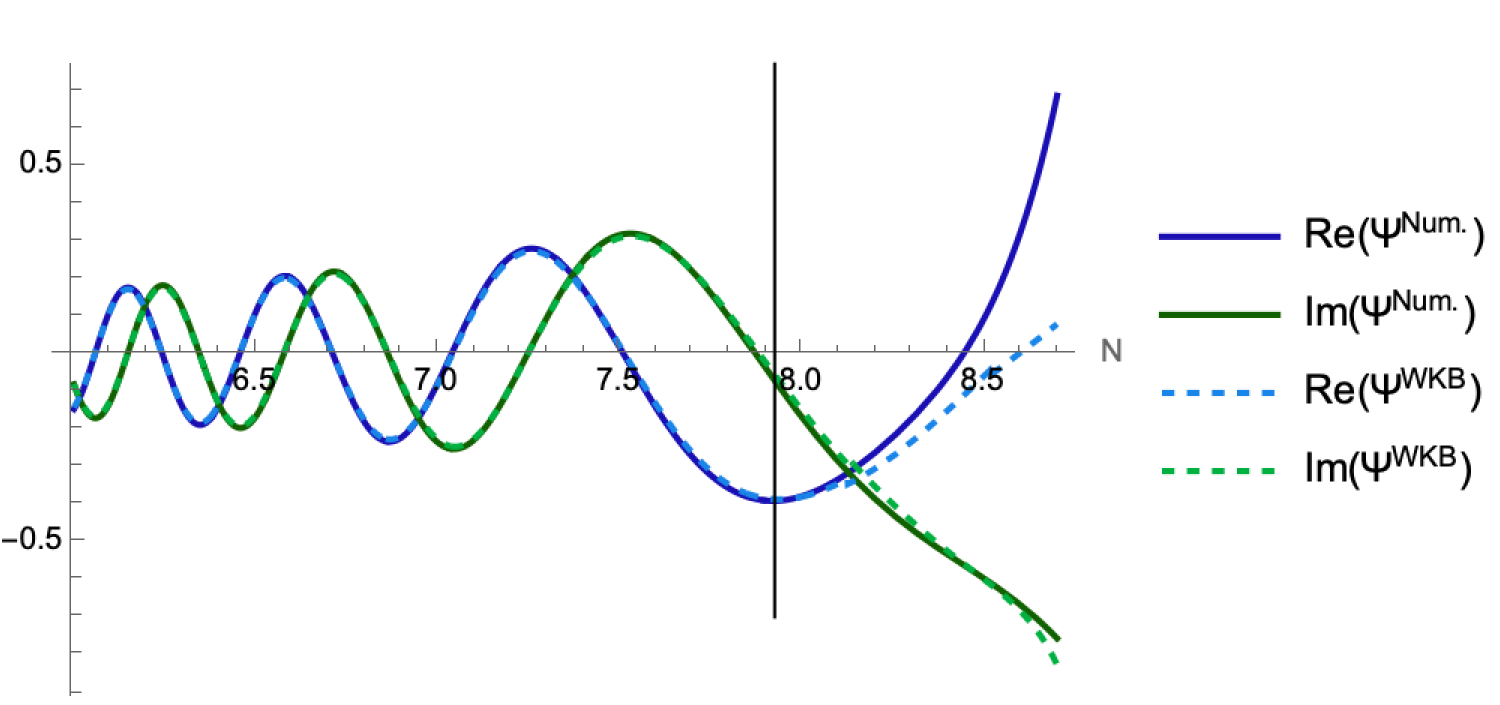}
    \caption{In these plots we consider the values $\epsilon=10^{-4}$, $\gamma=\eta=0$, $A_\alpha=10^{-7}$, $A_{c_s}=0.7$, $\sigma=1$, $N^*=5$, and $k/H=10^{4}$. LHS: Plot of the time evolution of $\alpha$ and $c_s^2$. The vertical black lines are placed at $N^*\pm 3 \sigma$. RHS: Comparison of WKB and numerical solutions to the equation of motion in Eq.~\eqref{eq:eomNnofric}. The WKB approximation shows the result using the leading order phase shift. The black line marks the breakdown of the WKB approximation at $N_\text{break}=0.9 N_{W_k=0}$ where $N_{W_k=0}=8.8$ is the e-fold where $W_k=0$.}
    \label{fig:WKBvsNum}
\end{figure}

In the first scenario, we consider a simple parametrization in which only one parameter controls the shape of the transition. We will work with a Gaussian parametrization of $\alpha$ and $c_s$ given by
\begin{subequations}
\begin{align}
    c_s^2(N)=&1- A_{c_s} e^{-\frac{(N-N^*)^2}{2 \sigma^2}} \ , \\
    \alpha(N)=& A_\alpha e^{-\frac{(N-N^*)^2}{2 \sigma^2}} \ ,
\end{align}
\label{eq:gaussians}
\end{subequations}
where we define the width of the transition as
\begin{equation}
    \wid= 6 \sigma \ .
\end{equation}
The parameters $A_{c_s}$ and $A_\alpha$ control the minimum and maximum of $c_s$ and $\alpha$ respectively,
\be
c^2_s(N^*)=1-A_{c_s} \ , \quad \alpha(N^*)=A_\alpha \ .
\ee
In Figure~\ref{fig:WKBvsNum}, we can see a comparison of the WKB approximation, used to obtain the phase shift, and the numerical solution for $\Psi_k$ as well as the time evolution of $\alpha$ and $c_s$. 

\begin{figure}[!h]
    \centering
    \includegraphics[width=\linewidth]{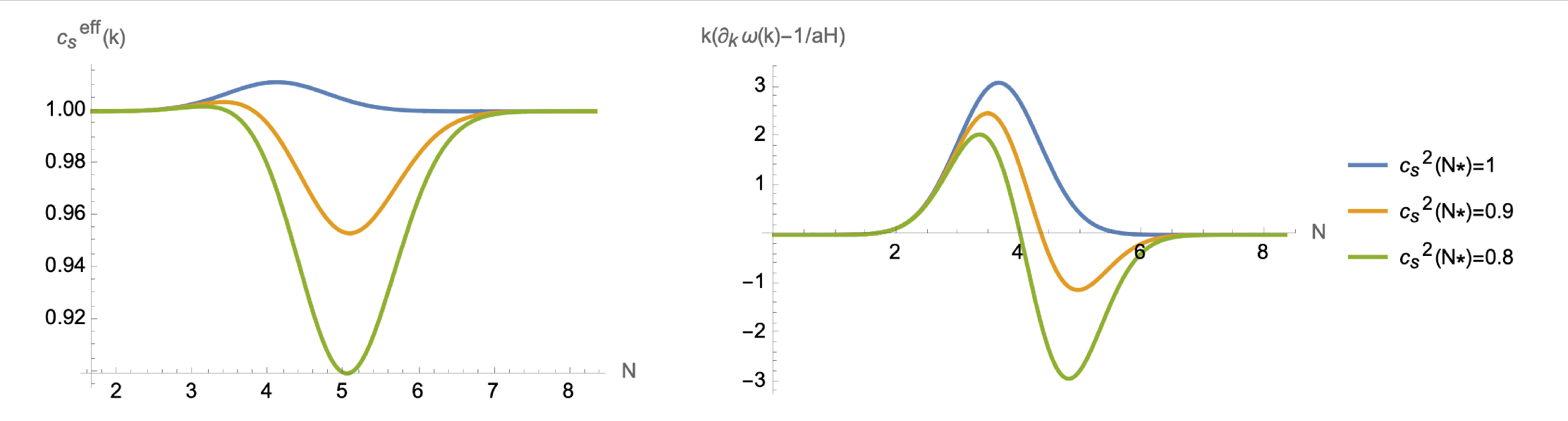}
    \caption{On the LHS we see the plot for the effective sound speed with Eq.~\eqref{eq:gaussians} for varying values of $c_s^2(N^*)=1-A_{c_s}$ and fixed $A_\alpha=10^{-5}$,  $N^*=5$, $\sigma=2/3$, and $\epsilon=10^{-4}, \eta=\gamma=0$. The RHS shows the plot for the integrand in the causality bound as written in Eq.~\eqref{eq:cboundIntegrand} for the same set of parameters.}
    \label{fig:cseff_integrand}
\end{figure}

To understand the bounds, we start by looking at the effective sound speed from Eq.~\eqref{eq:dispersion}
\begin{equation*}
   ( c_s^\text{eff}(k,N))^2=c_s^2(N)+\alpha(N) \frac{ k^2}{a^2H^2} \ ,
    \label{eq:effcs}
\end{equation*}
from which we see that if $c_s^2=1$ and $\alpha \neq 0$, then the effective sounds speed becomes superluminal, since $\alpha$ is positive. While this superluminality might only occur for a short period of time, this effect will be enhanced in the integrand of the spatial shift where the relevant quantity is $\omega\sim(k/H)c_s^\text{eff} E^N$, see Fig.~\ref{fig:cseff_integrand}. Even though the $\alpha$ contribution is suppressed by the EFT (see Eq.~\eqref{alphacond}), this term can dominate in the integrand at early times and integrate to a resolvable positive spatial shift. 

\begin{figure}[!h]
    \centering
    \includegraphics[width=0.45\linewidth]{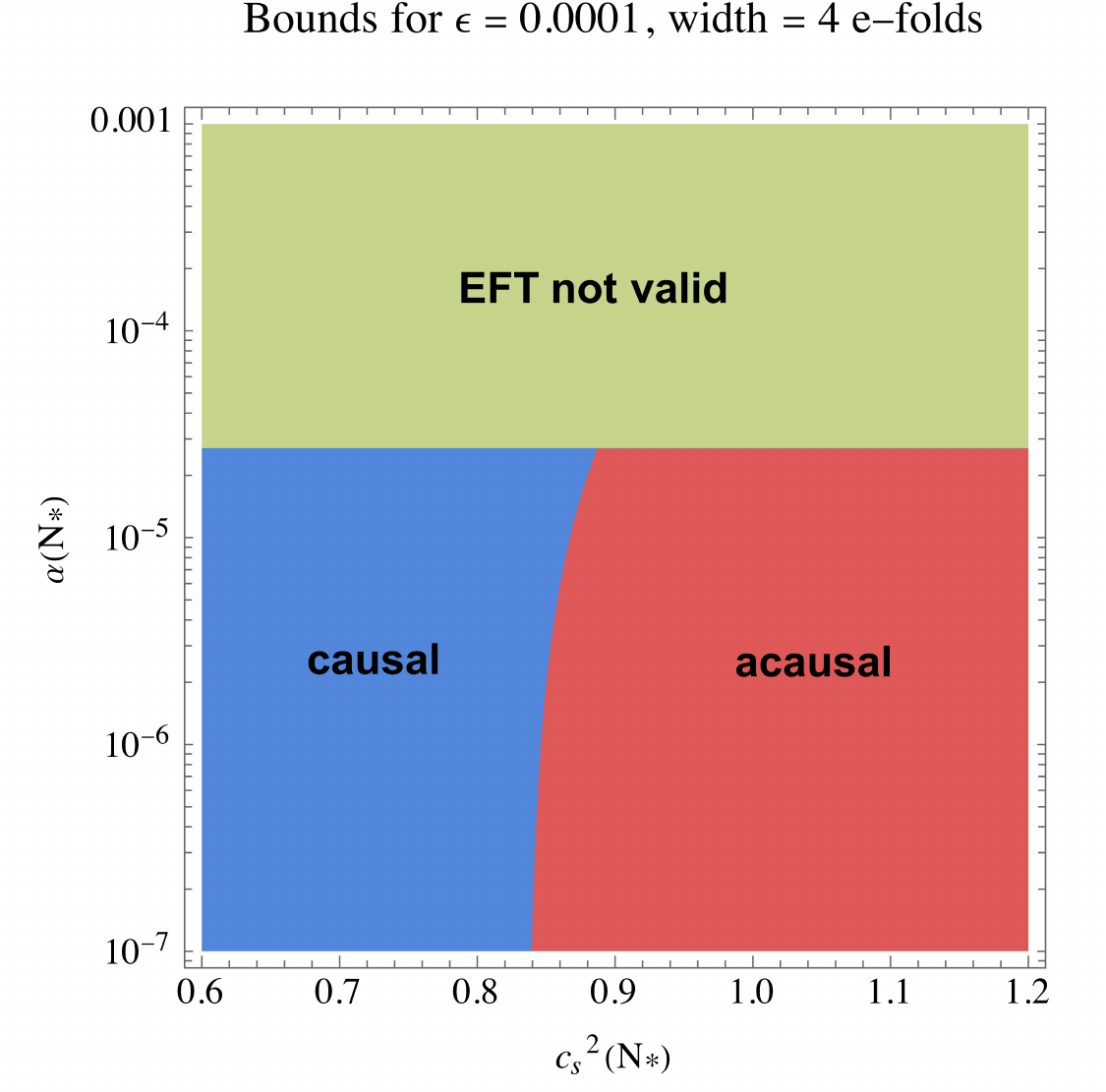}
    \hspace{0.05\linewidth}
    \includegraphics[width=0.45\linewidth]{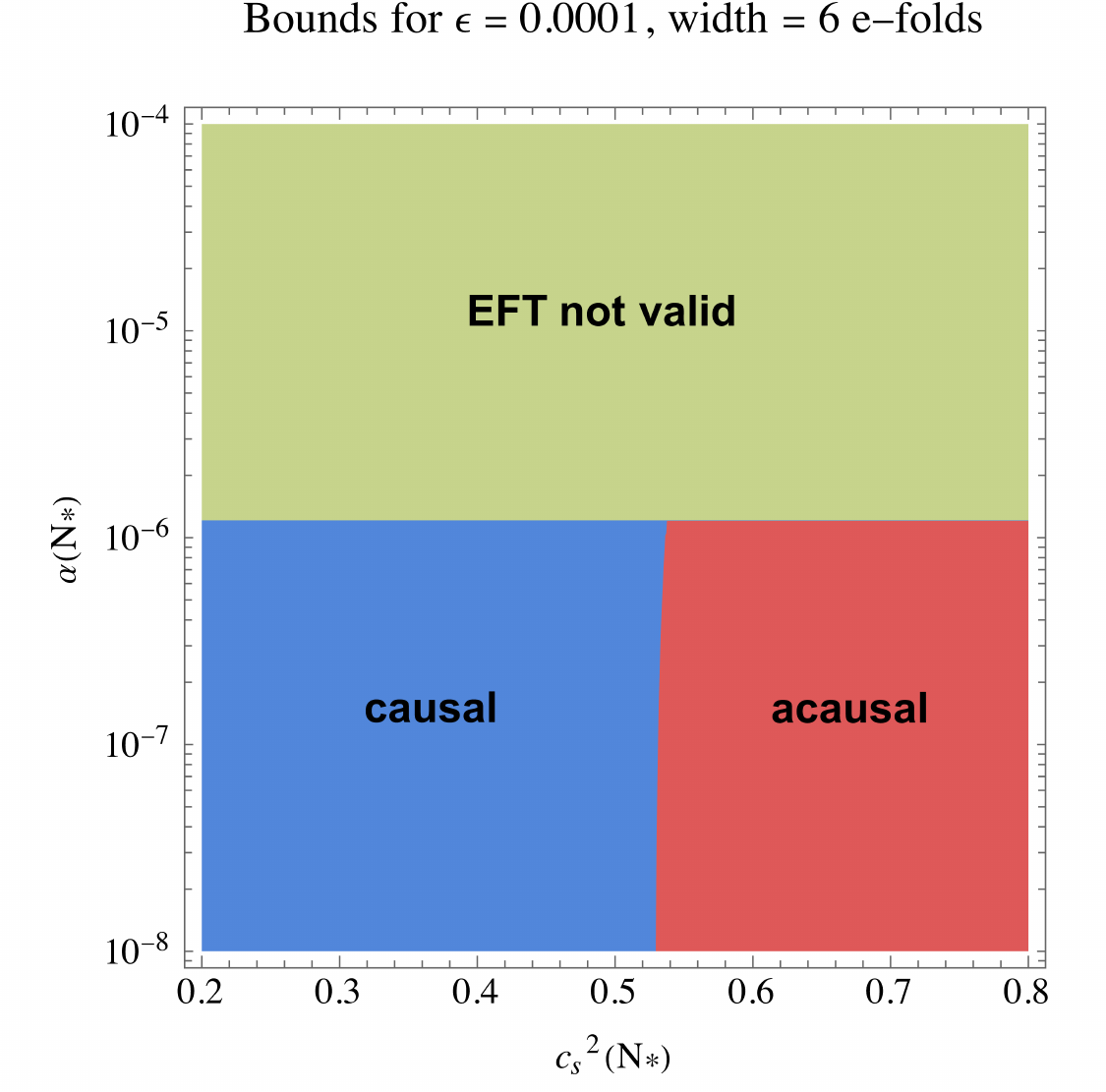}
    \caption{Bounds on the $\alpha-c_s^2$ plane for different widths of the transition with $\epsilon=10^{-4}, \eta=\gamma=0$. We see that a luminal speed of sound leads to acausal propagation. This can be avoided if the width is small enough, but in that case, the EFT is no longer valid. The green region corresponds to the values of $\alpha$ and $c_s^2$ for which Eq.~\eqref{alphacond} is violated in the integration region, that is, the modes that are within the horizon before the transition is over cannot be described by the EFT.}
    \label{fig:boundalphacs}
\end{figure}

The larger the width of the transition, the longer the effective sound speed will be superluminal which can lead to acausalities. Naively, it would seem that considering very small widths will allow for $c_s\geq1$ without violating causality, but these small widths in this parametrization also lead to sharp transitions which break the EFT. In fact, Eq.~\eqref{eq:GoverH} tells us that
\begin{equation}
    \frac{\Gamma}{H}=\left|\frac{\max{\frac{36 (N-N^*)e^{-18(N-N^*)^2/\wid^2}}{\wid^2}}}{A_{c_s}}\right|=\frac{6}{\sqrt{e} \ \wid}< 1  \quad \rightarrow \quad \wid>6/\sqrt{e}\sim 3.64\ .
\end{equation}
When the width of the transition satisfies this requirement, we find that luminal velocities lead to a resolvable, positive spatial shift for any value of $\alpha$. Thus, as in the constant effective sound speed scenario, if $c_s=1$ then $\alpha$ must vanish, and the same will apply to any higher order EFT operator contributing to the two-point function. In Fig.~\ref{fig:boundalphacs}, we show the bounds on the $\alpha-c_s^2$ plane where we use Eq.~\eqref{eq:gaussians} with $N^*=5$, $\epsilon=10^{-4}, \eta=\gamma=0$. The bounds are weakly dependent on $\alpha$, and for larger values of $\alpha$, larger values of $c_s^2$ are allowed. This can be understood from the fact that increasing $\alpha$ can break the validity of the EFT before the transition ends unless we lower $k/H$, which in turn makes the overall phase shift smaller. Hence, for larger $\alpha$, larger values of $c_s$ can lead to positive but unresolvable positive spatial shifts maintaining the theory causal. 

\begin{figure}[!hbt]
    \centering
    \begin{subfigure}[t]{0.48\textwidth}
        \centering
        \includegraphics[width=\linewidth]{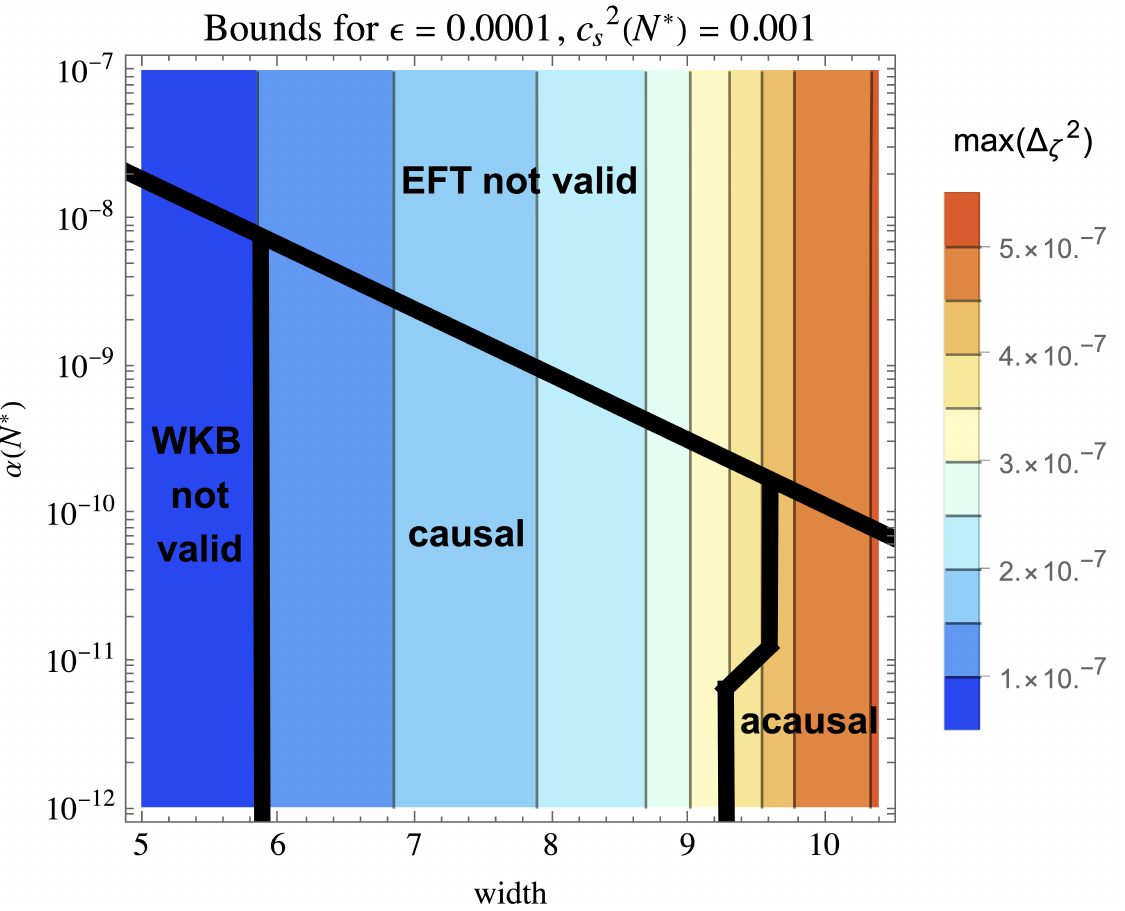}
    \end{subfigure}
    \hfill
    \begin{subfigure}[t]{0.48\textwidth}
        \centering
       \raisebox{0.5cm}{ \includegraphics[width=1\linewidth]{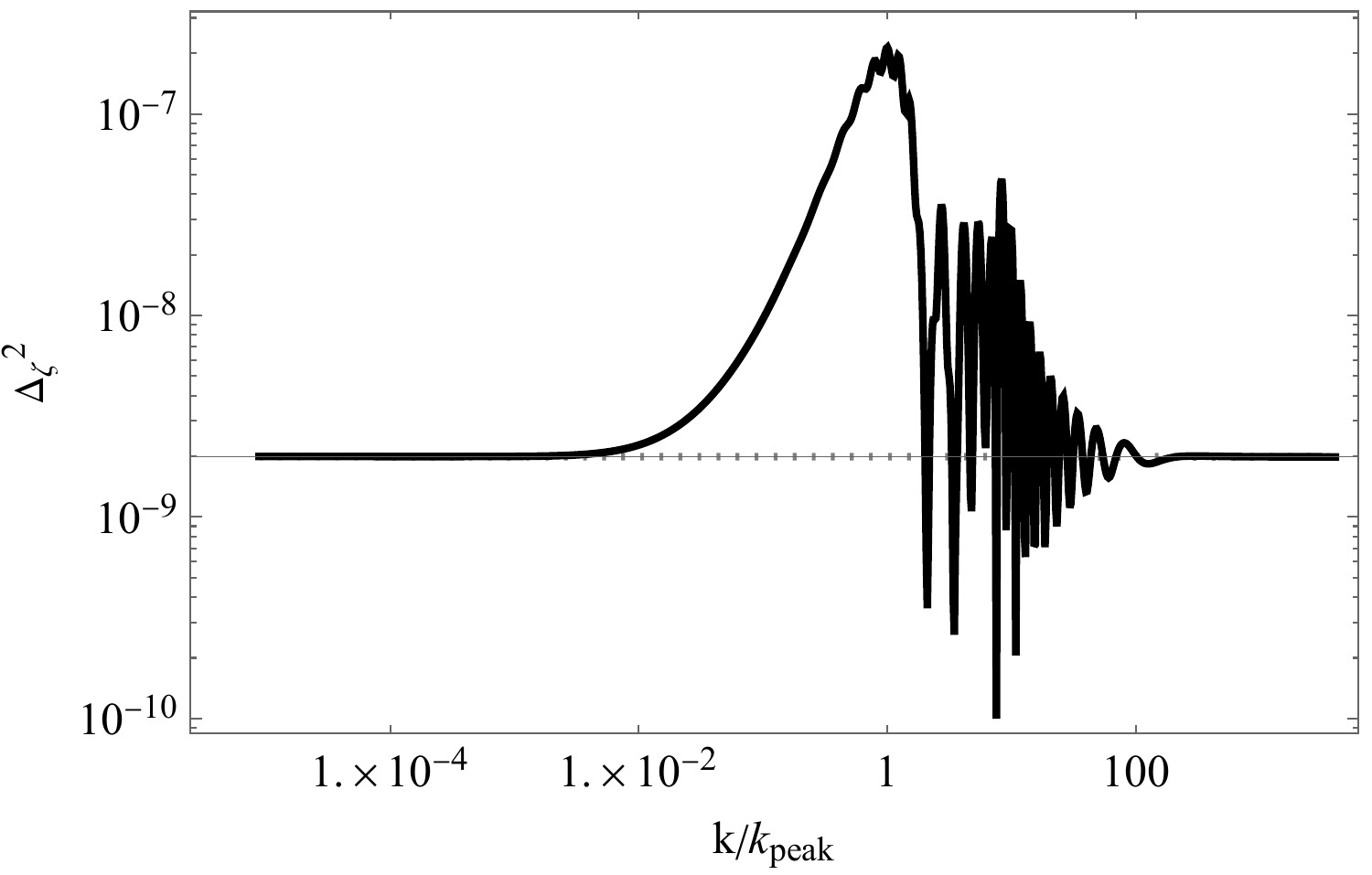}}
    \end{subfigure}
    \caption{LHS: Bounds on the $\alpha-$width plane for $c_s^2(N^*)=0.001$. The contours show the maximum value of the power spectrum, which has been normalized to $2 \times 10^{-9}$ at the initial integration e-fold. We observe that as we increase the width, the maximum of the power spectrum grows, but around 9 e-folds, the theory becomes acausal and $\max{(\Delta_\zeta^2)}\lesssim 4 \times 10^{-7}$ for a causal theory. 
    RHS: Plot of the power spectrum for different values of momentum normalized to the scale where the power spectrum peaks. The free parameters have the same values as in the LHS plot and additionally, $\alpha=10^{-10}$ and $\wid=7$.}
    \label{fig:boundalphawidth}
\end{figure}

We can also look at the bounds in the $\alpha-\wid$ plane which will allow us to put upper bounds on the maximum power spectrum. At fixed $A_{c_s}$, the power spectrum grows as we increase the width of the transition. This means that the time for which the effective sound speed can be superluminal is increased and consequently the integrand becomes positive for more e-folds. Thus, as we increase the width, we make the spatial shift grow until, at some point, it is positive and resolvable. The bounds we find are shown in Fig.~\ref{fig:boundalphawidth}. We find that $\max{(\Delta_\zeta^2)}\lesssim  10^{-7}$ for a causal theory. Note that taking $c_s^2(N^*)=0$ will not increase the power spectrum significantly but will cause the breaking of the WKB approximation in all the parameter space since $W_k$ will cross zero before the end of the transition. This situation already happens for small widths in the case shown in Fig.~\ref{fig:boundalphawidth}. Reducing the value of $\epsilon$ will move the line showing the breaking of the WKB approximation to the left of the plot, that is, to smaller widths.

In this Gaussian parametrization even if we choose a large width and a small $c_s$, which increases the power spectrum, we will not obtain very large values of the power spectrum since $c_s$ is near its minimum for a very small time. Similarly, even if the $\alpha$ term dominates at the transition the fact that it is only near its extrema for a short period makes the approximation in Eq.~\eqref{PS_ST} invalid and the power spectrum does not grow as $\alpha$ is decreased. To get a larger growth of the power spectrum, we consider two different parametrizations in the next subsections that will allow $c_s$ to stay near the minimum for a longer time\footnote{We note that in \cite{Ballesteros:2018wlw}, a parametrization of the form $1-c_s^2,\alpha\sim \text{exp}(\tanh((N-N^*)/\sigma)$ was considered. This allows the parameters to spend a longer period of time in their extrema, and increase the power spectrum more than in our Gaussian example while having a single parameter controlling the shape of the transition. It also avoids issues with ill-defined initial conditions. Similar to the Gaussian case, the growth in this example will be bounded by causality.}.

%%%%%%%%%%%%%%%%%%%%%
\subsection{Hyperbolic tangent parametrization}
\begin{figure}[!ht]
    \centering
    \includegraphics[width=0.48\linewidth]{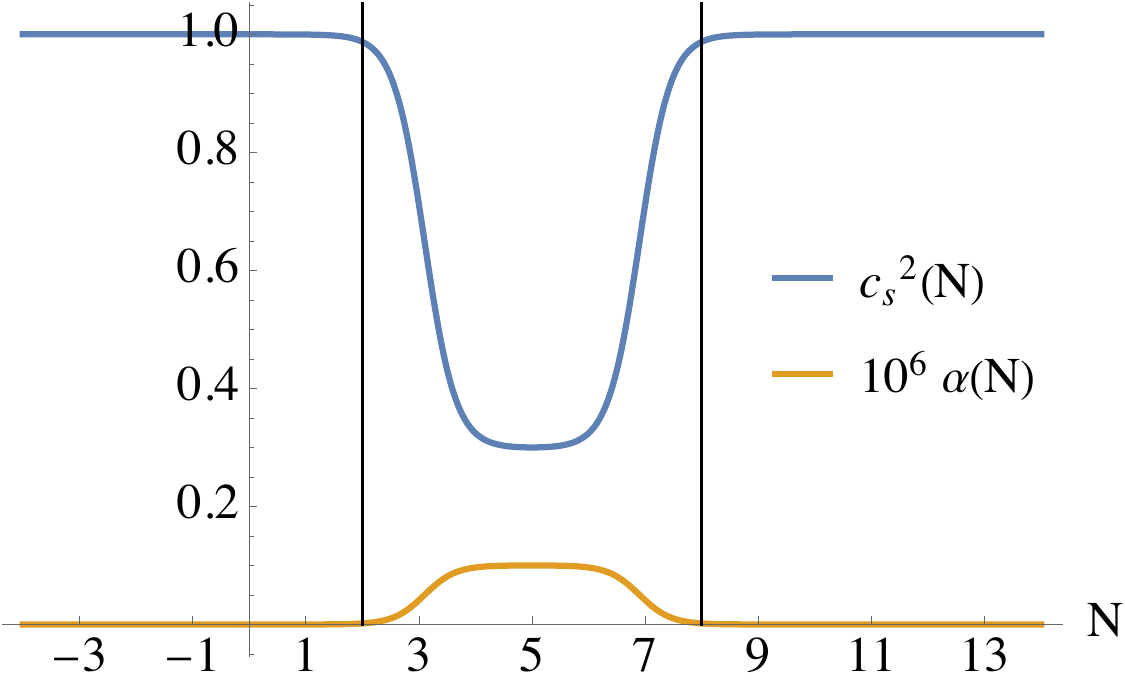}
    \includegraphics[width=0.48\linewidth]{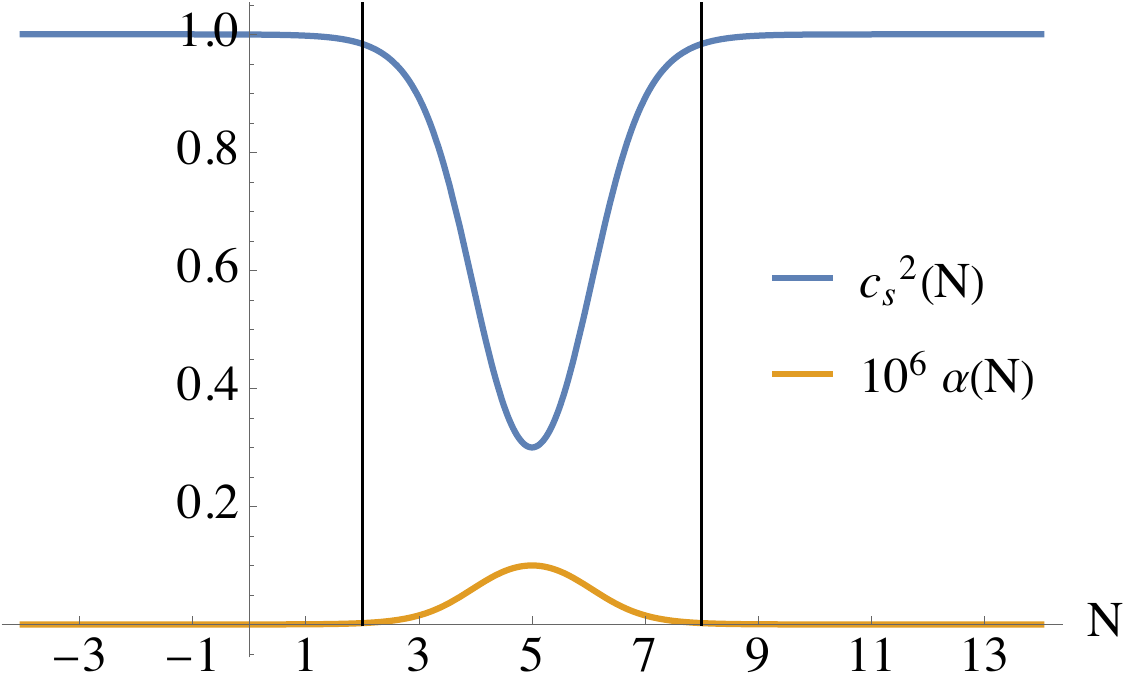}
    \caption{Plots of the time evolution of $\alpha$ and $c_s^2$ for the values $\epsilon=10^{-4}$, $\gamma=\eta=0$, $A_\alpha=10^{-7}$, $A_{c_s}=0.7$, $\wid=6$, and $N^*=5$. The LHS has $\sigma=0.55$ and the RHS $\sigma=1$.}
    \label{fig:plotalphacs_tanh}
\end{figure}

In the previous parametrization, the growth of the power spectrum was at most two orders of magnitude. This is due to the Gaussian profile selected which allows the value of $c_s$ to stay near its minimum for a very short e-fold length. To get a larger growth of the power spectrum, we now consider a parametrization in terms of a hyperbolic tangent as follows
\begin{subequations}
\begin{align}
    c_s^2(N)=&1 - A_{c_s} \frac{\left( 1 + 
    \tanh\left(N-N_{\star}-\frac{\wid-4\sigma }{2\sigma}\right)
    \tanh\left(N_{\star}-N-\frac{\wid-4\sigma }{2\sigma}\right)
    \right)}{
    1 + \tanh^2\left( \frac{\wid-4\sigma }{2\sigma} \right)
} \ , \\
    \alpha(N)=& A_\alpha \frac{\left( 1 + 
    \tanh\left(N-N_{\star}-\frac{\wid-4\sigma }{2\sigma}\right)
    \tanh\left(N_{\star}-N-\frac{\wid-4\sigma }{2\sigma}\right)
    \right)}{
    1 + \tanh^2\left( \frac{\wid-4\sigma }{2\sigma} \right)
} \ ,
\end{align}
\label{eq:tanh}
\end{subequations}
which gives rise to the profiles observed in Fig.~\ref{fig:plotalphacs_tanh}. In this parametrization, $\sigma$ controls the sharpness of the transition, the parameter width controls the total width of the transition when width $>2$ otherwise the transition's width is a more complicated function for both width and $\sigma$. As before, we can analyse when the transition is too fast such that $\Gamma/H>1$ and the EFT breaks, this happens for $\sigma<0.5$ when width$>2$ as observed in Fig.~\ref{fig:boundstanh}.

\begin{figure}[!ht]
    \centering
    \includegraphics[width=0.42\linewidth]{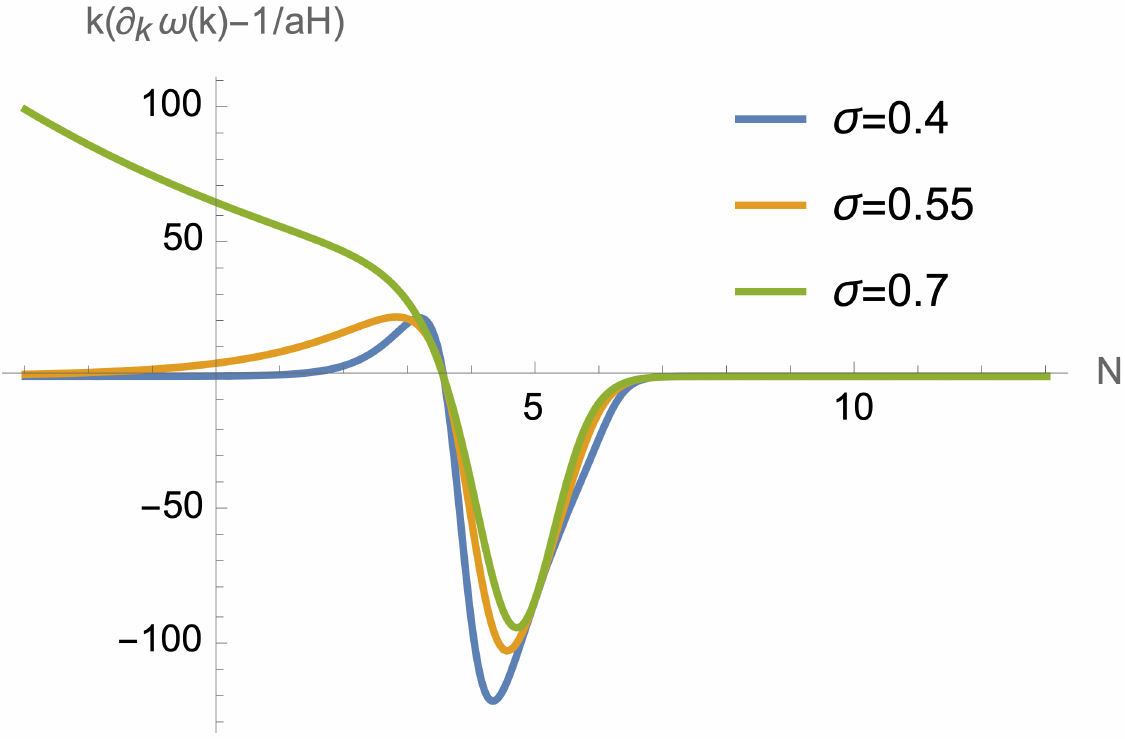}
    \caption{Plot of the integrand of the causality bound in Eq.~\eqref{eq:cboundIntegrand} for the parametrization in Eq.~\eqref{eq:tanh} with fixed $A_\alpha=10^{-7}$,  $N^*=5$, $\wid=4$, $\epsilon=10^{-4}, \eta=\gamma=0$ and  varying $\sigma$. The blue line gives $\Gamma/H>1$, the yellow one is within the regime of validity of the EFT and allows for a Bunch Davies vacuum, while the green one has $\lim_{\tau\rightarrow-\infty}c_s^\text{eff}>1$ and makes the integrand blow up in the past.}
    \label{fig:integrandtanh}
\end{figure}

When working with this parameterisation, we have to consider slow enough transitions, large $\sigma$, so that $\Gamma/H<1$ but also fast enough such that we can define the Bunch-Davies vacuum in the far past. If $\sigma$ is too large, the transition is very slow and the $\alpha$ term in the effective sounds speed doesn't shut down in the past. In other words, we have to ensure that $\lim_{\tau\rightarrow-\infty}c_s^\text{eff}=1$. If this is not the case, even for a very small deviation from one, the integrand of the spatial delay will blow up because the scattering-like problem is not well defined, see Fig.~\ref{fig:integrandtanh}. 

\begin{figure}[!ht]
    \centering
    \includegraphics[width=0.485\linewidth]{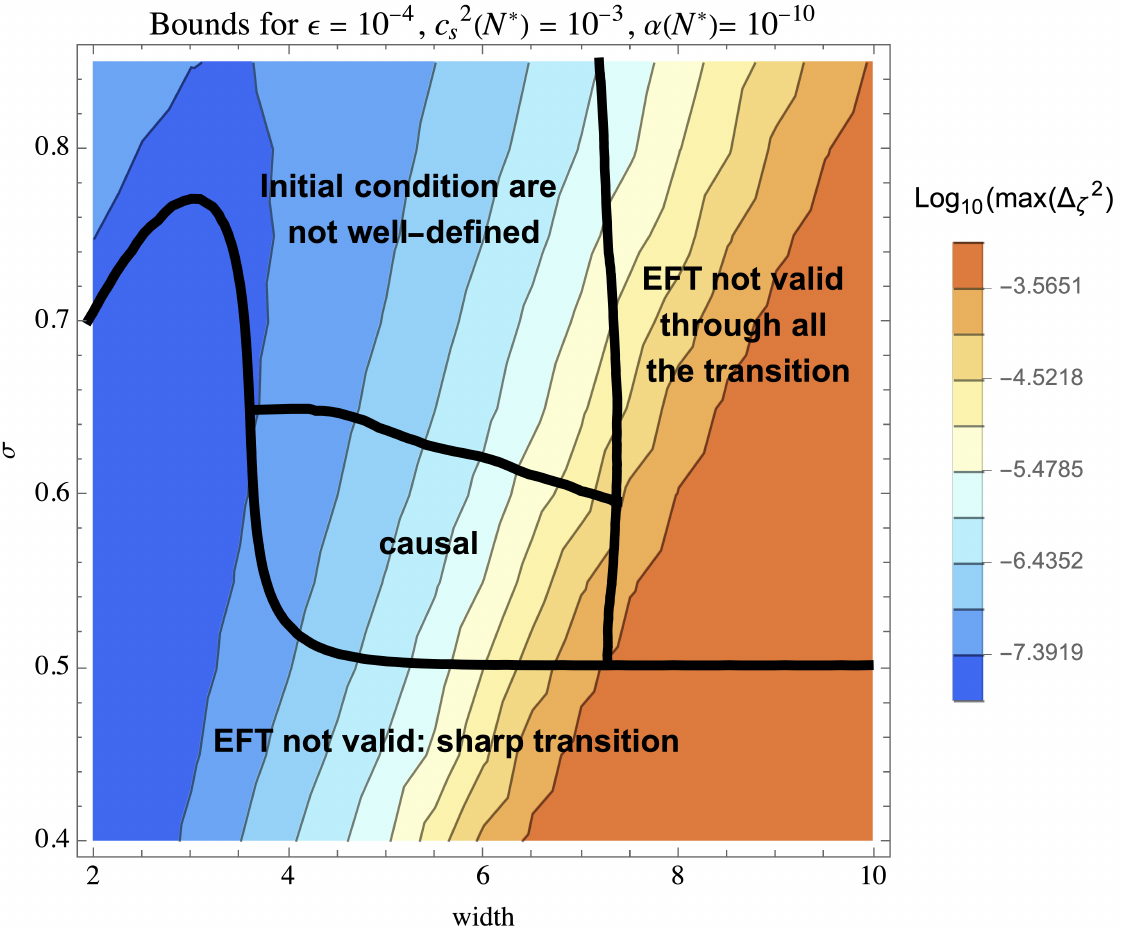}
    \includegraphics[width=0.48\linewidth]{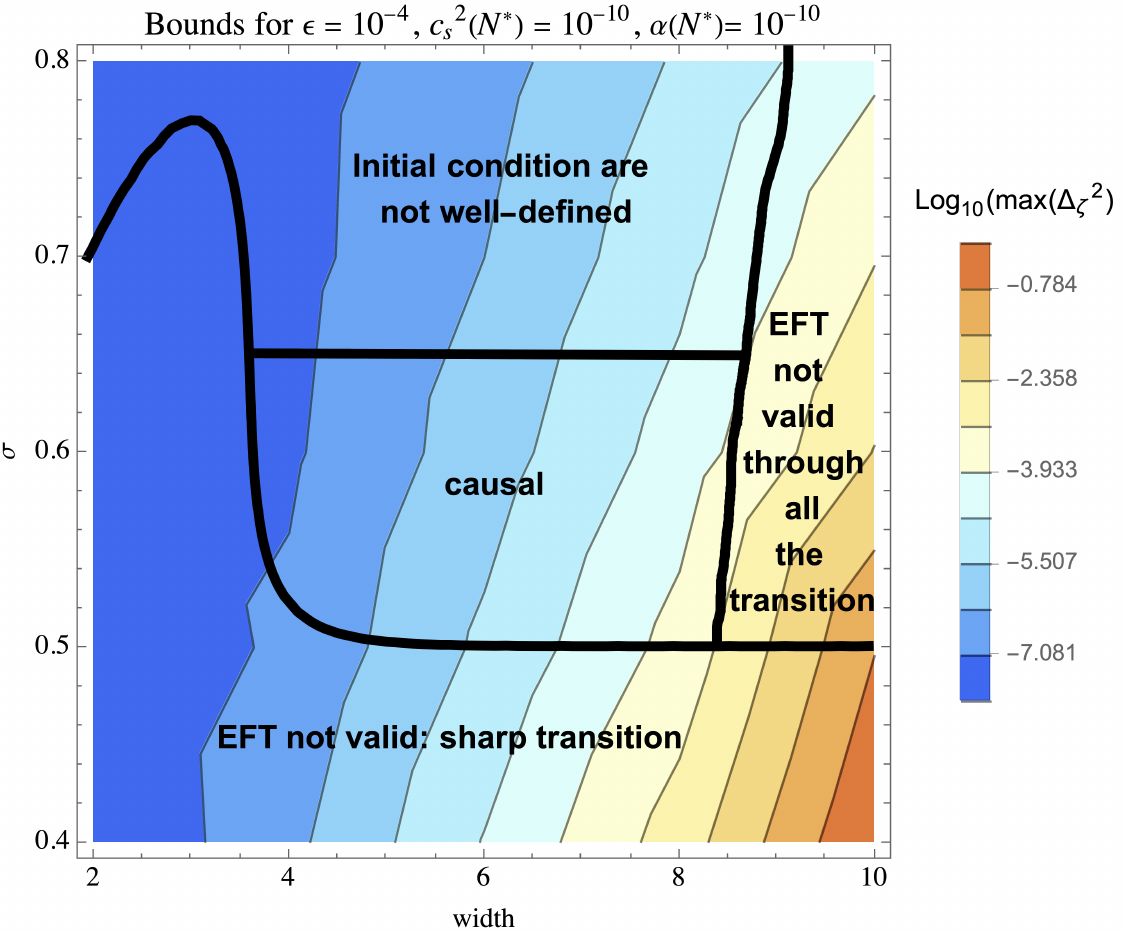}
    \caption{Bounds on the maximum of the power spectrum in the $\wid$ and $\sigma$ plane. At small $\sigma$, the EFT breaks down due to a sharp transition, that is, $\Gamma/H>1$. As mentioned in the text, for $\wid<2$, the width of the transition depends on both $\wid$ and $\sigma$ which causes the complicated dependence of the EFT breaking line at low widths. For large $\sigma$, the $\alpha$ term in the effective sound speed blows up in the past; the effect of this is seen in Fig.~\ref{fig:integrandtanh}. Last, for large widths, the EFT breaks down before the end of the transition for modes that stay within the horizon. As we increase $\alpha$, this boundary moves towards smaller widths and as we decrease it, the boundary moves towards larger widths. The LHS has $c_s(N^*)=10^{-3}$ while the RHS $c_s(N^*)=10^{-10}$. The EFT and perturbative expansions were set up so that in the LHS the $c_s$ term always dominates and in the RHS the $\alpha$ term dominates near $N^*$.}
    \label{fig:boundstanh}
\end{figure}

\begin{figure}[!ht]
    \centering
    \includegraphics[width=0.48\linewidth]{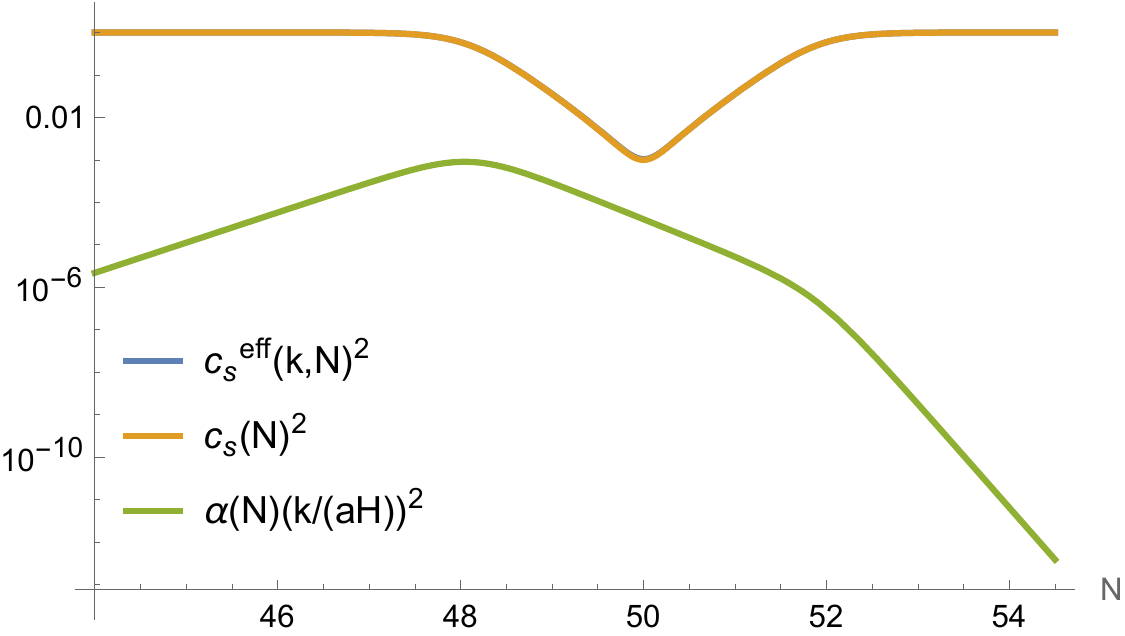}
    \includegraphics[width=0.48\linewidth]{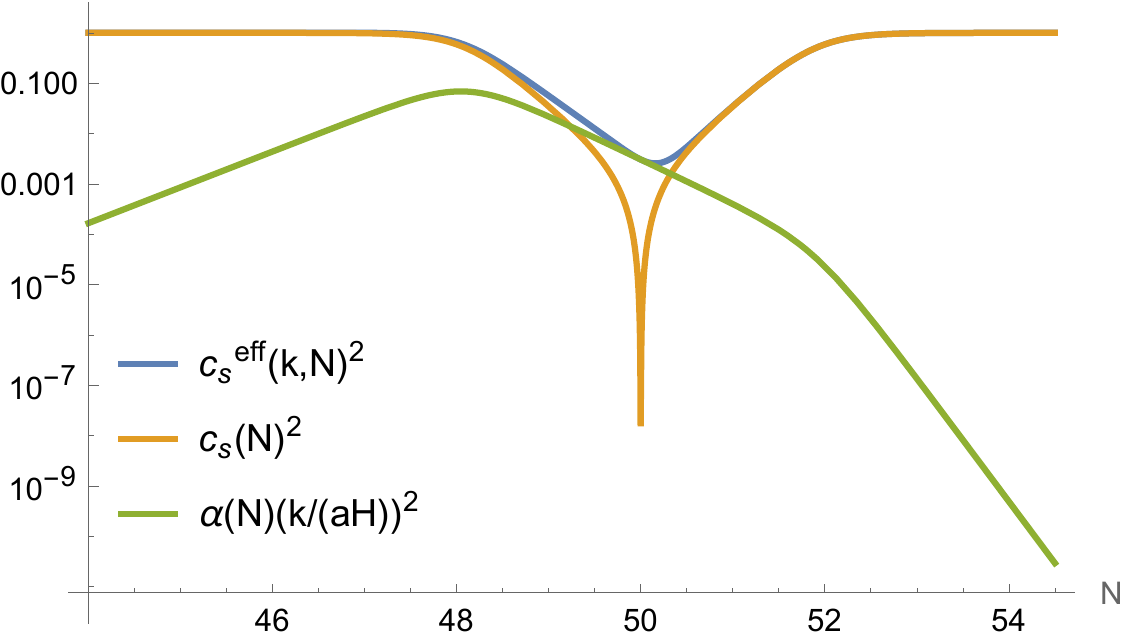}
    \caption{Plots of the effective sounds speed and the separate contributions from the leading order term ($c_s(N)$) and the next-to-leading order ($\alpha(N)$) contribution in derivatives for fix $A_\alpha=10^{-10}$,  $N^*=50$, $\wid=6$, $\sigma=0.55$, and $\epsilon=10^{-4}, \eta=\gamma=0$. On the LHS we set $A_{c_s}=10^{-3}$ and the $\alpha$ term never dominates while in the RHS we have $A_{c_s}=10^{-10}$ and the $\alpha$ term dominates during the transition. }
    \label{fig:csefftanh}
\end{figure}

\begin{figure}[!ht]
    \centering
    \includegraphics[width=0.5\linewidth]{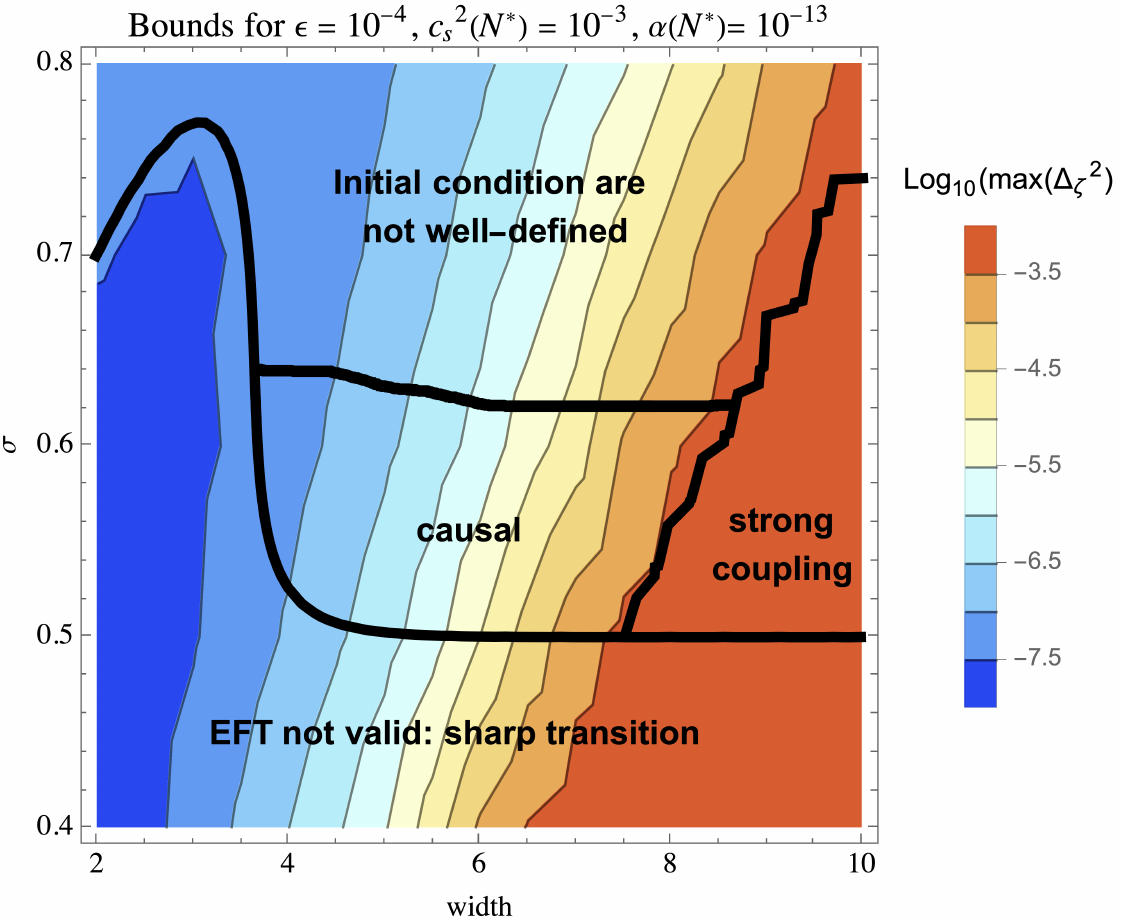}
    \caption{Bound on the maximum of the power spectrum in the $\wid$ and $\sigma$ plane with the same parameters as in the LHS of Fig.~\ref{fig:boundstanh} but with $\alpha=10^{-13}$ which extends the validity of the EFT to longer lasting transitions. In this case, while the EFT is valid, the well-known constraint on increasing the power spectrum comes from entering the strong coupling regime. The strong coupling region here corresponds to the parameter space where $2\pi^2 c_s^4(N) < 10 \Delta^2_\zeta(N)$ at some e-fold.}
    \label{fig:boundstanhSC}
\end{figure}

In this parametrization, the growth of the power spectrum is not bounded by causality, but by the validity of the EFT and the strong coupling regime, as observed in Fig.~\ref{fig:boundstanh} and \ref{fig:boundstanhSC}. In fact, we observe that causality is rarely violated in this model since the parameters that would lead to violations of causality, such as large $\sigma$ or $\wid$, also lead to ill-defined initial conditions and breaking of the EFT respectively. A slow transition allows the $\alpha$ term to dominate in the past and leads to an effective sound speed that becomes larger than one for longer periods. This was observed in Fig.~\ref{fig:cseff_integrand} in the Gaussian parametrization and Fig.~\ref{fig:integrandtanh} in the current hyperbolic tangent case. The difference is that the Gaussian shuts down the transition fast enough in the past to have well-defined initial conditions while the hyperbolic tangent does so only for small enough $\sigma$.

We find that at large enough $\alpha$, only a small island of parameter space is allowed by the validity of the EFT, well-defined initial conditions, and causality. Furthermore, for $c_s(N^*)\gtrsim10^{-3}$ and $\alpha(N^*)\gtrsim10^{-8}$ no region of parameter space remains viable. For smaller $\alpha$, while the EFT remains valid, small values of $c_s$ that lead to the growth of the power spectrum also lead to strong coupling. Thus, as is well known, if the leading order $c_s$ term dominates the power spectrum cannot grow without entering the strong coupling regime where the tree-level perturbative results used are no longer valid, see Fig.~\eqref{fig:boundstanhSC}. If $c_s$ decreases drastically during the transition, the sub-horizon modes will remain within the validity of the EFT for longer-lasting transitions. In these scenarios, the higher derivative ($\alpha$) term dominates during the transition and avoids the strong coupling problems arising from small $c_s$. Similar to the case of $c_s$ dominating, here too there will be no region of parameter space that remains viable if $c_s(N^*)\gtrsim10^{-10}$ and $\alpha(N^*)\gtrsim10^{-5}$.

\begin{figure}[!ht]
    \centering
    \includegraphics[width=0.6\linewidth]{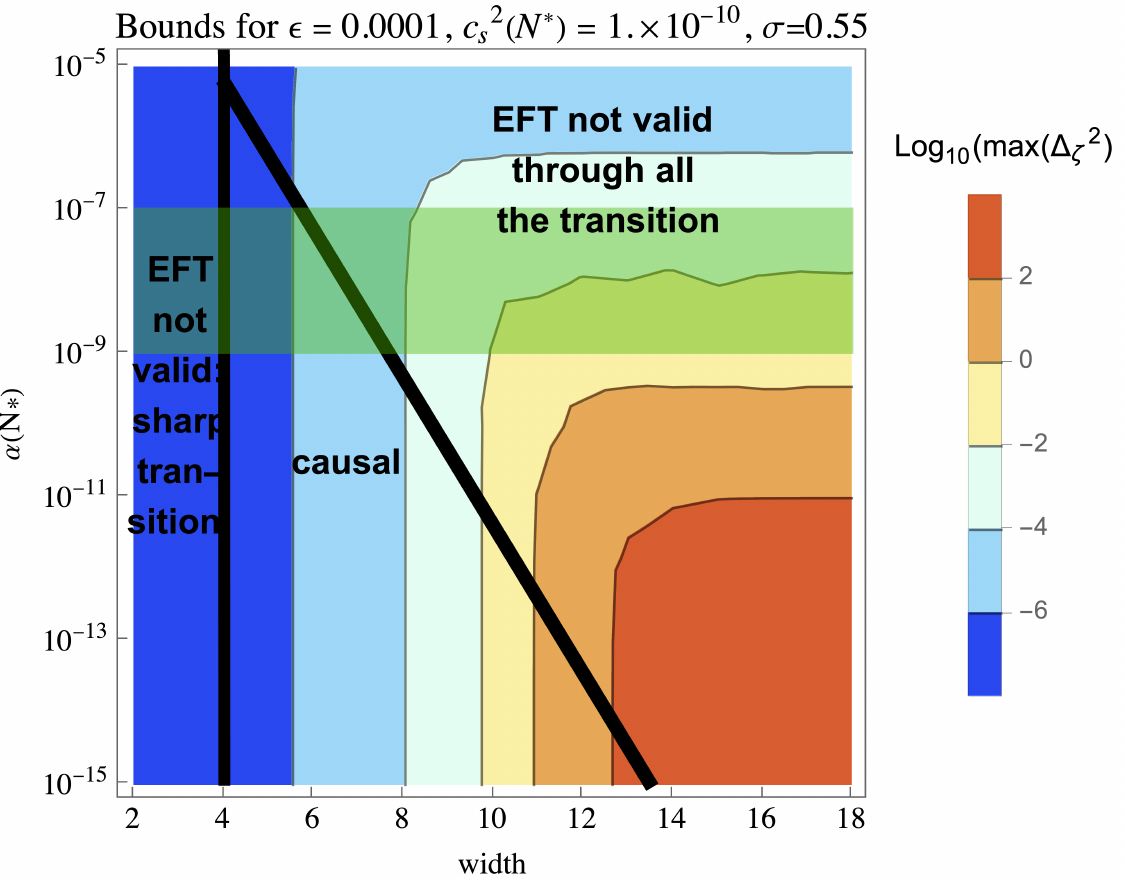}
    \caption{Contours of the maximum of the power spectrum for different values of $\alpha$ and $\wid$ with fix $A_{c_s}=10^{-10}$,  $N^*=50$, $\sigma=0.55$, and $\epsilon=10^{-4}, \eta=\gamma=0$. We can observe that while the maximum grows as we decrease $\alpha$, it eventually saturates. Furthermore, in the region where it depends on $\alpha$, the modes that are within the horizon are not within the regime of validity of the EFT during the whole transition.}
    \label{fig:PStanh}
\end{figure}

We also note that in the regime where the $\alpha$ term dominates during the transition, smaller $\alpha$ and larger widths lead, as expected, to an increase of the power spectrum; see Fig.~\ref{fig:PStanh}. While very small values of $\alpha$ seem to allow for a large growth of the power spectrum, these would correspond to a very fine-tuned situation. For example, fine-tuning $\alpha$ to be of order $\lesssim 10^{-10}$ but allowing higher-order operators to have their natural values would lead to inconsistencies since those operators should have been included in our analysis as they would dominate over $\alpha$. More importantly, such operators will give quantum corrections to $\alpha$, meaning that such small values are not radiatively stable.

%%%%%%%%%%%%%%%%%%%%%
\subsection{Split Gaussian parametrization}
\begin{figure}[!ht]
    \centering
    \includegraphics[width=0.48\linewidth]{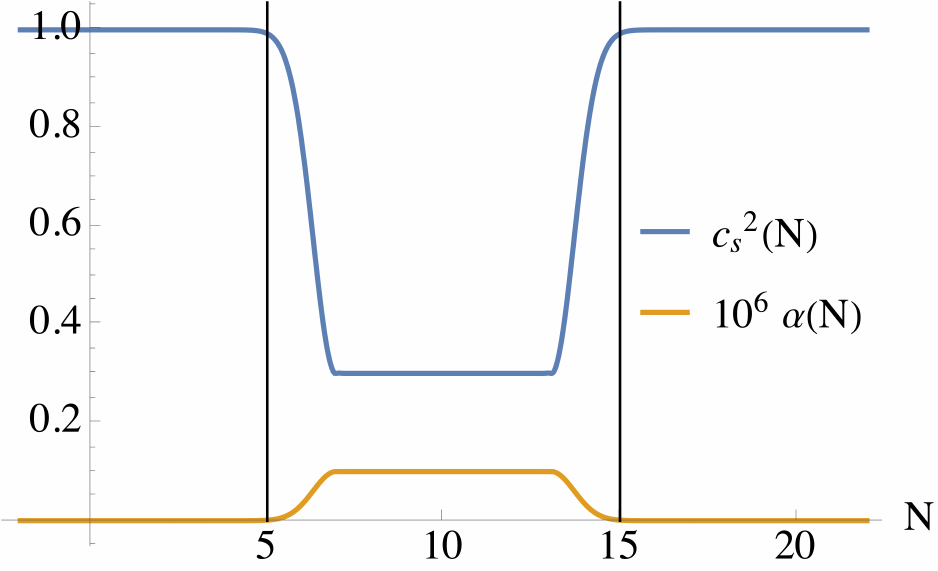}
     \includegraphics[width=0.48\linewidth]{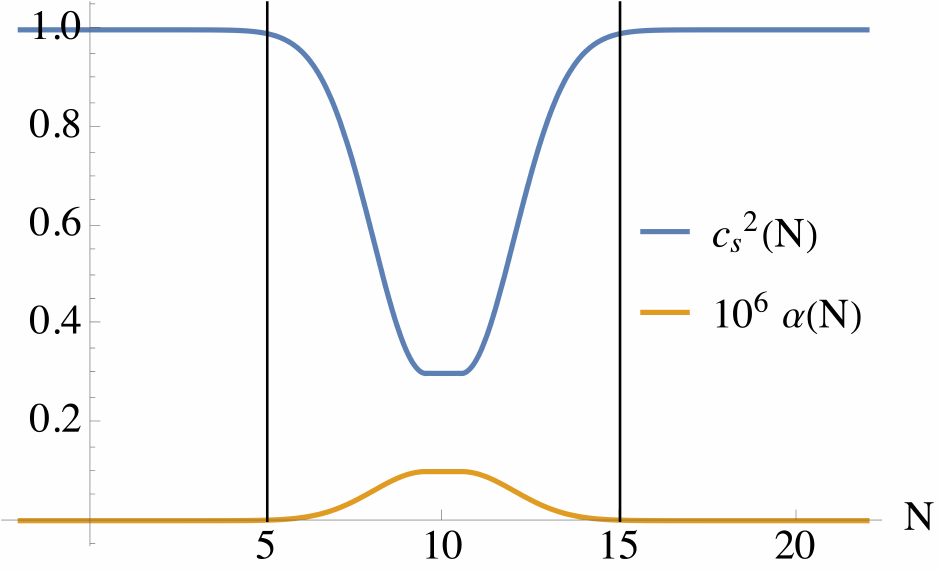}
    \caption{Plots of the time evolution of $\alpha$ and $c_s^2$ in the parametrization of Eq.~\eqref{eq:splitgaussian} for the values $\epsilon=10^{-4}$, $\gamma=\eta=0$, $A_\alpha=10^{-7}$, $A_{c_s}=0.7$, $\wid=10$, and $N^*=10$. The LHS has $\sigma=0.65$ and the RHS $\sigma=1.5$ .}
    \label{fig:plotalphacs_splitgauss}
\end{figure}

In this last parametrization, we completely disentangle the time that $c_s$ and $\alpha$ spend at their extrema from the sharpness of the transition and consider Gaussian transitions that always allow for well-defined initial transitions to understand which characteristics lead to breaking of causality and which ones to breaking of the EFT without having initial condition problems. The parametrization is as follows
\begin{subequations}
\begin{align}
    c_s^2(N) &= 1 - A_{c_s} \Bigg[
    \bigg( \Theta_{\text{sm.}}\left(-N + \left(N^* + \frac{w}{2}\right)\right)
    + \Theta_{\text{sm.}}\left(N - \left(N^* - \frac{w}{2}\right)\right) - 1 \bigg) \nonumber \\
    &\qquad + e^{-\frac{\left(N - \left(N^* - \frac{w}{2}\right)\right)^2}{2 \sigma^2}} 
    \Theta_{\text{sm.}}\left(-N + \left(N^* - \frac{w}{2}\right)\right) \nonumber \\
    &\qquad + e^{-\frac{\left(N - \left(N^* + \frac{w}{2}\right)\right)^2}{2 C_C^2}} 
    \Theta_{\text{sm.}}\left(N - \left(N^* + \frac{w}{2}\right)\right)
    \Bigg], \\
    \alpha(N) &= A_\alpha \Bigg[
    \bigg( \Theta_{\text{sm.}}\left(-N + \left(N^* + \frac{w}{2}\right)\right)
    + \Theta_{\text{sm.}}\left(N - \left(N^* - \frac{w}{2}\right)\right) - 1 \bigg) \nonumber \\
    &\qquad + e^{-\frac{\left(N - \left(N^* - \frac{w}{2}\right)\right)^2}{2 \sigma^2}} 
    \Theta_{\text{sm.}}\left(-N + \left(N^* - \frac{w}{2}\right)\right) \nonumber \\
    &\qquad + e^{-\frac{\left(N - \left(N^* + \frac{w}{2}\right)\right)^2}{2 C_C^2}} 
    \Theta_{\text{sm.}}\left(N - \left(N^* + \frac{w}{2}\right)\right)
    \Bigg],
    \label{eq:splitgaussian}
\end{align}
\label{eq:gaussian_split}
\end{subequations}
where $\Theta_{\text{sm.}}$ is a smoothed version of the Heaviside theta function written in terms of Gauss error functions as
\be
\Theta_{\text{sm.}}(N-N')=\frac{1}{2}\left(1+ \text{erf}\left(\frac{N-N'}{0.1}\right)\right) \ .
\ee
Since the transition speed is given by the Gaussian, we already know from the analysis in the subsection \ref{subsec:gauss} that $\Gamma/H>1$ if $\sigma<1/\sqrt{e}$. The parameter $w$ controls the number of e-folds that the functions stay at their extrema and the width is given by
\begin{equation}
    \wid=w+6 \sigma \ .
\end{equation}
Examples of the shape of the transition can be observed in Fig.~\ref{fig:plotalphacs_splitgauss} for a fast and a slow transition.

\begin{figure}[!ht]
    \centering
    \includegraphics[width=0.48\linewidth]{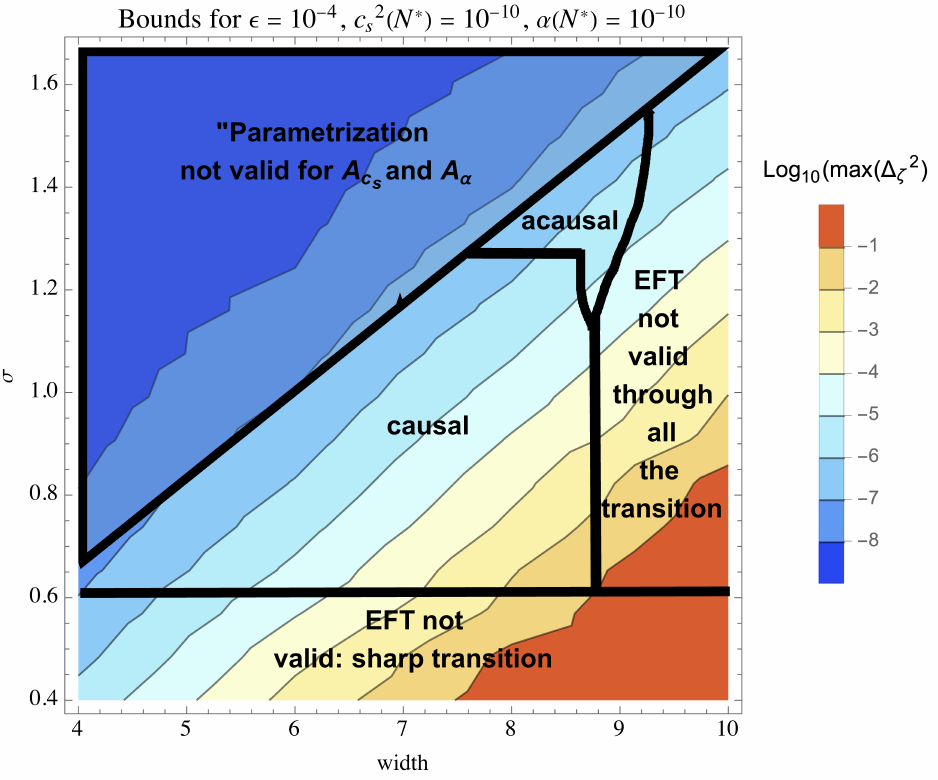}
    \includegraphics[width=0.5\linewidth]{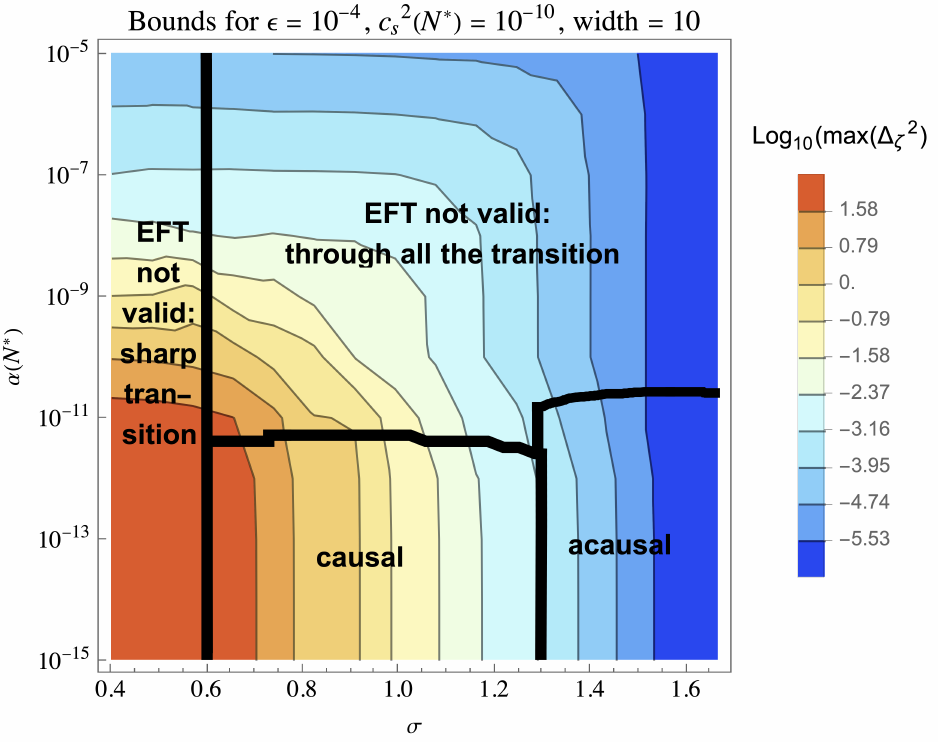}
\caption{Causal and causal regions in the $\wid$ and $\sigma$ and $\alpha(N^*)$ and $\sigma$ planes. In the upper left triangle of the LHS figure, the parametrization that gives the extrema of $c_s$ and $\alpha$ in terms of $A_{c_s}$ and $A_\alpha$ is no longer valid. For $\sigma<1/\sqrt{e}$ we have that $\Gamma/H>1$ and for larger $\sigma$, that is, slower transitions, the propagation becomes causal. Larger $\alpha$ can lead to modes inside the horizon not being in the regime of validity of the EFT for the whole transition.}
    \label{fig:boundsSplitGauss}
\end{figure}

In Fig.~\ref{fig:boundsSplitGauss}, we observe the regions of acausal propagation in the $\wid-\sigma$ plane. Now that these two variables are completely disentangled, we see that the speed of the transition is what controls the violations of causality. If the transition is too slow, the $\alpha$ term grows in the past, leading to a superluminal effective sound speed. In the upper left triangle of Fig.~\ref{fig:boundsSplitGauss}, we have $\wid< 6 \sigma$ which makes the smoothed theta functions in the parametrization in Eq.~\eqref{eq:splitgaussian} to overlap such that the relations $c_s(N^*)=1-A_{c_s}$ and $\alpha(N^*)=A_\alpha$ no longer hold and we have instead $c_s(N^*)>1-A_{c_s}$ and $\alpha(N^*)<A_\alpha$.

%%%%%%%%%%%%%%%%%%%%%
\subsection{Summary of general features}
Now that we have analyzed different parametrizations for the transitions of the effective sound speed, we can summarize the main features that we observed. 

\paragraph{(Super)luminal $c_s$ leads to a free theory}
Whenever we consider $c_s(N^*)\geq1$ and $\alpha(N^*)>0$, there is always a mode with momentum $k$ which is within the horizon and remains within the validity of the EFT during the whole transition. This means that the effective sound speed will become superluminal during the whole transition. Furthermore, we can not have arbitrarily short-length transitions since this would require fast transitions that lead to $\Gamma/H>1$ and break the EFT. Thus, the effect always integrates to a positive resolvable spatial shift and the mode propagates in an acausal manner. The only way to avoid this is if $\alpha$ and all other higher derivative corrections to the two-point function and interacting terms vanish. Terms contributing to higher-point functions at tree-level will generate loop corrections to the two-point function and no matter how small they are, this will lead to a mode with acausal propagation. Note that our finding—that a luminal speed of sound in a causal EFT requires the theory to be free—is consistent with the observations in \cite{Baumann:2015nta,Baumann:2019ghk}.

\paragraph{Acausality from slow transitions}
If we consider a model with the fewest free parameters where the transition's speed and the time the parameters spend at their extrema are related, larger times at the extrema / slower transitions will give rise to a growth of the $\alpha$ term at early times where it can dominate and lead to a superluminal effective speed of sound. If this superluminality lasts for long enough it will integrate to a positive and resolvable spatial delay that breaks causality. On the other hand, if we allow for more free parameters and disentangle the time spent at the extrema from the speed of the transition, we see that only the speed of the transition is relevant and is responsible for the superluminal features of the effective sound speed. This becomes clear when looking at Fig.~\ref{fig:boundsSplitGauss}.

\paragraph{Bounds on the growth of the power spectrum}
The maximum value of the power spectrum increases as we consider longer transitions. When the length of time that the parameters spend at their extrema and the speed of the transition are entangled, as in the Gaussian parametrization, considering longer transitions leads to acasual propagation. Thus, in these scenarios, one can constrain the growth on the power spectrum by requiring causal propagation. Otherwise, causality does not lead to any bounds since acausal propagation arises for large $\sigma$, but the maximum of the power spectrum decreases slightly as we slow down the transition. In these cases, the bound on the growth of the power spectrum comes from the validity of the EFT or the strong coupling regime. The strong coupling bound for $c_s\rightarrow 0$ can be avoided if the $\alpha$ term dominates during the transition. Despite this, we find that regardless of $c_s$ or $\alpha$ dominating during the transition, there is a very fine-tuned (or vanishing) region of parameter space that is allowed for natural values of $\alpha$. 

%%%%%%%%%%%%%%%%%%%%%%
\section{Causality bounds on a UV completion}
\label{sec:2field}
We have seen that the EFT permits regions of parameter space that violate causality bounds when propagating over certain backgrounds, even though the effective action remains within its regime of validity. This raises the question of whether such causality violations can occur when the effective theory is derived from a UV completion, for instance, by integrating out heavy degrees of freedom. If all the parameters of the EFT arise from a causal UV theory, we will show that the EFT naturally protects itself, ensuring that its validity inherently enforces causality across any background.
Let us consider a model of two scalar fields with an action given by,
\begin{align}
    S=S_{\mathrm{EH}}+\int d^4x\sqrt{-g}(\gamma_{ab}(\phi)\partial_\mu\phi^a\partial^\mu\phi^b+V(\phi^a)) \ ,
\end{align}
where $S_{\mathrm{EH}}$ is the Einstein-Hilbert action, $\gamma_{ab}$ the field metric on target space and $V(\phi^a)$ the inflationary potential. We will assume that there is a homogeneous inflationary solution $\phi_0^a(t)$ that breaks Lorentz invariance and that there are two fields. The action for the curvature field $\zeta$ and the isocurvature field $\psi$ is obtained by decomposing the fluctuations in terms of the normal and orthogonal directions defined by the background inflationary solution $\phi_0^a(t)$ (For more details see \cite{GrootNibbelink:2001qt,Achucarro:2010da,Achucarro:2012yr}.)

At second order, the action for $\zeta$ and $\psi$ is given by,
\begin{align}
    S_{(2)}=\frac{M_{\mathrm{pl}}}{2}\int d^4 x a^3\left[2\epsilon\zeta^2-\frac{2\epsilon}{a^2}(\partial_i\zeta)^2+\dot\psi^2-\frac{1}{a^2}(\partial_i\psi)^2-M^2\psi^2-4\Omega \frac{\dot\phi_0}{HM_{\mathrm{pl}}}\dot\zeta\psi\right] \ , \label{eq:2field}
\end{align}
where $\Omega$ is the coupling between the two fields and $M$ is the  mass of the isocurvature field $\psi$.
Notice that only the tangential direction is slow roll suppressed whereas the normal direction is a coupled massive field. We are interested in obtaining an EFT after integrating out the field $\psi$  when the effective mass $M^2\gg H^2$. 

We can integrate out the second field by assuming that time derivatives are suppressed with respect to the other terms in the equation of motion. Replacing the equation of motion for $\psi$ into the action yields the non-local action,
\begin{align}
    S=\frac{M_{\mathrm{Pl}^2}}{2}\int d^4xa^3 2\epsilon \left[\left(\frac{M^2c_s^{-2}+k^2/a^2}{M^2+k^2/a^2}\dot\zeta^2-\frac{k^2}{a^2}\zeta^2\right)\right] \ ,
    \label{eq:nonLocalAction}
\end{align}
where we the speed of sound $ c_s^{-2}=1+\tfrac{4\Omega^2}{M^2} $, is luminal only when the two fields are decoupled. This action can be made local by expanding in inverse powers of the mass $M$. The action above assumes energies below a cut-off scale $\Lambda_\mathrm{UV}$, above which the two fields need to be considered. This energy scale can be obtained by considering the superhorizon limit of the theory in which the modes are characterised by two frequencies $\omega_\pm$ given by,
\begin{align}
    \omega_{\pm}^2=\frac{M^2}{2c_s^2}+\frac{k^2}{a^2}\pm\frac{M^2}{2c_s^2}\sqrt{1+\frac{4c_s^2 k^2(1-c_s^2)}{a^2M^2}} \ .
\end{align}
The single field EFT is valid when $\omega_+\gg \omega_-$.  Notice that at leading order in $H/M$, the cut-off scale is given by $\Lambda_\mathrm{UV}^2=\omega_+^2=M^2c_s^{-2}$. 

\begin{figure}
    \centering
    \includegraphics[width=0.55\linewidth]{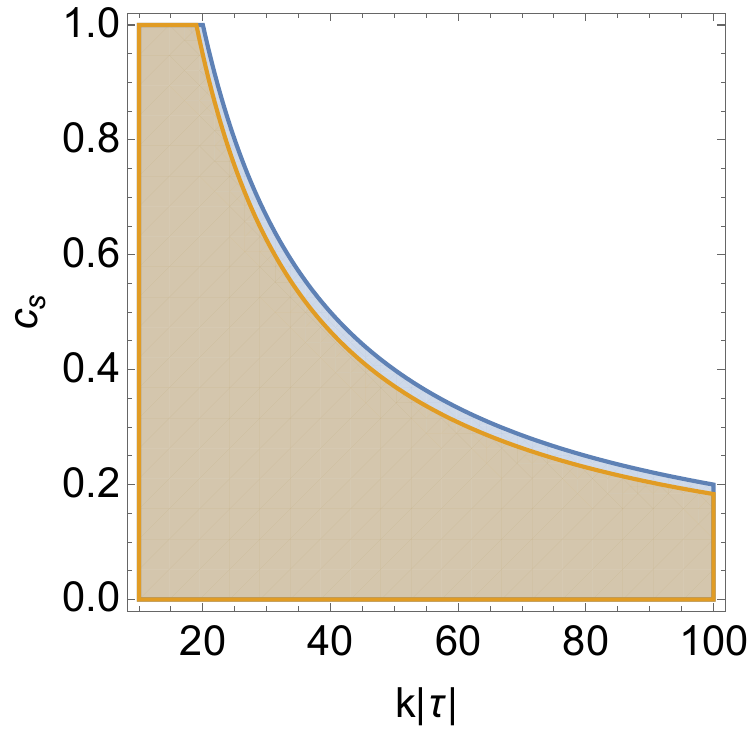}
    \caption{The plot shows the allowed regions from \eqref{eq:bound2field} (orange) and from the requirement that the physical momentum is below the cutoff (blue). The parameters are $\tau_i=\tau=-1$, $\tau_f=-0.1$, $M=20 H$, and we only consider modes that remain within the horizon, i.e., $k\vert\tau_f\vert\gg1$. Notice that the causality constraint is always satisfied as long as we remain within the regime of validity of the single field EFT. At large $k\vert\tau\vert$ the allowed values of $c_s$ tend to zero.}
    \label{fig:Causality2fieldmodel}
\end{figure}

In order to study the causality constraints on the effective action \eqref{eq:nonLocalAction} we can solve the equation of motion using the WKB approximation and expanding in powers of $H/M$. As before, we can analyse first the simplest case of constant sound speed and $\alpha$ in a de Sitter background. Thus, the constraint in \eqref{eq:cbound} becomes,
\begin{align}
       \frac{\tau_{\mathrm{final}}\left(c_s^2+c_s^2(1-c_s^2)\frac{M^2}{H^2} k^2\tau_{\mathrm{final}}^2\right)^{1/2}-\tau_{\mathrm{ini}}(c_s^2+c_s^2(1-c_s^2)\frac{M^2}{H^2}  k^2\tau_{\mathrm{ini}}^2)^{1/2}}{\tau_{\mathrm{final}}-\tau_{\mathrm{ini}}}\leq 1 \ .
       \label{eq:bound2field}
\end{align}
In the plot in Fig.(\ref{fig:Causality2fieldmodel}) we find that the causality constraints are satisfied as long the energy of the theory is below the cut-off $\Lambda_\mathrm{UV}$, as expected since the UV theory was causal.

When considering time-dependent Wilson coefficients and hence time-dependent effective speeds of sound, we will again conclude that the validity of the EFT protects the theory from acausal propagation, a well-known feature in generic covariant theories. 

In order to have acausal propagation, we need the higher-order in $k$ contribution in the effective sound speed dominates at early times. This requires that $k/(aH)\gg M/H$ which together with the validity of the EFT, $c_s (k/(aH)) (H/M)\ll 1$, imply that $c_s\ll1$. Thus in the regions where $c_s\ll1$ and the higher-order in $k$ contribution can dominate, the effective sound speed does not become superluminal. After exploring the whole parameter space numerically for different parametrizations of the transition we can further confirm that this is correct and, as expected, the EFT arising from a known UV completion does not have acausal models thanks to the EFT protection.

%%%%%%%%%%%%%%%%%%%%%%
\section{Discussion and conclusions}
\label{sec:Conclusions}
In this paper, we have analysed the constraints that arise on the effective sound speed of the comoving curvature whose two-point function gives the primordial power spectrum. We considered a scenario where the effective sound speed undergoes a transition such that the power spectrum can grow at scales smaller than the CMB where we have tight observational constraints. To understand the effects of different features of the transition, we looked at three different parametrizations that allowed us to draw general conclusions. One of the interesting conclusions that we reached is that if $c_s=1$, causality bounds require the theory to be free. This is consistent with observations made in \cite{Baumann:2015nta,Baumann:2019ghk}. We also noted that causality violations can arise even when $c_s\rightarrow 0$ during the transition if the transition is very slow. This happens since higher derivative terms grow in the past as $k/(aH)$, thus if they are not shut down fast enough, they can grow drastically and lead to a superluminal effective sound speed for a long enough period of time such that the spatial shift becomes resolvable. 

We also concluded that obtaining a large growth of the power spectrum is challenging for natural values of the EFT parameters. Notably, if only one parameter controls the speed and length of the transition, causal propagation gives a bound on the maximum growth of the power spectrum. If these two features of the transition are disentangled and controlled by different parameters, the growth of the power spectrum is bounded due to the validity of the EFT, the strong coupling regime, and the naturalness of the EFT parameters. 

There are several possible extensions of this work. In principle, it may be possible to study causality bounds in models of primordial black holes (PBHs) that undergo a phase of ultra-slow-roll (USR) evolution. If we simply consider a phenomenological parametrization of the potential such that the perturbation's speed of sound only changes due to the curvature of the spacetime, we would not expect any causality bounds on this theory. The reason is that the phase shift would only have contributions from an effective potential. In \cite{CarrilloGonzalez:2023emp}, it was shown that these scenarios do not lead to resolvable spatial shifts as expected. On the other hand, if we consider a well-defined EFT with USR background evolution that includes higher-order derivative terms properly parametrised, then one could find bounds on these terms and their effects. Additionally, in this work, we have only considered modifications to the dispersion relation involving polynomial in $k$ terms, but more general dispersion relations could also be of interest, particularly in the context of scalar-induced gravitational waves~\cite{deRham:2018red,Baker:2022eiz}.

Beyond the two-point function, it would be valuable to investigate the impact of causality on nonlinear terms. One approach to this would be to understand how to extend our method to study higher order connected retarded function  $G_c^{(3)}$ (see \cite{AguiSalcedo:2023nds,Baumgart:2020oby} for some recent progress on causality for higher order correlation functions), which could lead to interesting constraints on the parameter space of certain primordial non-Gaussianities. There are existing bounds obtained by requiring that the flat-space scattering amplitudes of the theory satisfy positivity conditions ~\cite{Baumann:2015nta, Grall:2020tqc, Grall:2021xxm} (similar approaches for other EFTs are found in \cite{Melville:2019wyy,deRham:2021fpu,Traykova:2021hbr,Kim:2019wjo,Herrero-Valea:2019hde,Ye:2019oxx}), and it would be compelling to compare these constraints to those derived from causality considerations.

%%%%%%%%%%%%%%%%%%%%%%
\section*{Acknowledgements}
We would like to thank Andrew Tolley for insightful discussions on causality. MCG thanks Lucien Heurtier for useful discussions regarding primordial black holes. SC is supported in part by the STFC Consolidated Grants ST/T000791/1 and ST/X000575/1 and by a Simons Investigator award 690508. MCG is supported by the Imperial College Research Fellowship.

%%%%%%%%%%%%%%%%%%%%%%
\appendix
\section{Method for obtaining causality bounds} \label{ap:method}
In order to obtain the causality bounds implied by Eq.~\eqref{eq:causal} we proceed as follows. We start by choosing the integration limits such that $\alpha(N),c_s^2(N)-1\rightarrow 0$ at $N_\text{ini}$ and $N_\text{final}$ and ensure that the integrand is smaller than $10^{-6}$ at those points. We require that the WKB approximation, Eq.~\eqref{eq:wkb}, holds and that the EFT is valid, Eq.~\eqref{alphacond} and \eqref{eq:GoverH}. This requires that
\begin{equation}
    \delta_\text{wkb} e^{N_\text{final}}<\frac{k}{H}< \epsilon_\text{eft} \left(\frac{e^N}{\sqrt{\alpha(N)}}\right)_{\text{min}} \ , \label{eq:rangek}
\end{equation}
where $\epsilon_\text{eft}\leq 0.5$, $\delta_\text{wkb}>2$, and the subscript min indicates the minimum within the integration bounds. The constraint in Eq.~\eqref{eq:GoverH} is analysed separately for each parametrization of the transition. We fix $\delta_\text{wkb}=2.5$, which sets the lower bound on the allowed momentum values. Then, we compute the spatial shift up to the following orders
\begin{align}
\left(\sqrt{\alpha}k/(aH)\right)^2\left(aH/k\right) \quad &\text{for $c_s$ always dominating $c_s^\text{eff}$} \ , \\
\left(c_s^2+\left(\sqrt{\alpha}k/(aH)\right)^2\right)\left(aH/k\right) \quad & \text{for $\alpha$ allowed to dominate $c_s^\text{eff}$} \ ,
\end{align}
and ensure that higher-order contributions in $\sqrt{\alpha}k/(aH)\ll1$ and $aH/k\ll1$ are indeed suppressed and their contribution to the spatial shift is $k\Delta r<0.05$; this gives the maximum allowed $\epsilon_\text{eft}$. Once $\epsilon_\text{eft}$ is fixed, we verify that the EFT is valid within our integration limits, that is, 
\begin{equation}
    N_\text{final} < \log\left(\frac{\epsilon_\text{eft} }{\delta_\text{wkb}}\left(\frac{e^N}{\sqrt{\alpha(N)}}\right)_{\text{min}}\right) \ .
\end{equation}
If this bound is not satisfied, the EFT breaks down before the transition in $\alpha$ and $c_s$ ends. We do not consider these cases since they don't give a well-defined scattering-like problem and give an ill-defined phase shift. In fact, one can see from Fig.~\ref{fig:cseff_integrand} that integrating only to an early e-fold in the middle of the transition could give the incorrect result of a positive, resolvable spatial shift. We also check that inside the integration region, we do not have any ``turning point'' at which $W_k=0$ and the WKB approximation breaks down. This type of behaviour would happen if we were to consider a transition to negative $c_s^2$ values or when both $c_s$ and $\alpha$ are too small. In such cases, one could do a standard WKB matching near the turning point to obtain the spatial shift at the fixed slice $N=N_\text{final}$. These scenarios are beyond the scope of this paper. 

Having ensured that the WKB approximation holds, the EFT is valid, and perturbation theory holds throughout the whole integration region, we proceed to compute the spatial shift for the range of $k/H$ in Eq.~\eqref{eq:rangek}. We do so for the three different parametrizations in the bulk of the paper while varying the parameters
\begin{equation}
    \{A_{c_s}, A_\alpha,\sigma, \wid \} \ ,
\end{equation}
that respectively control the amplitudes of $c_s$ and $\alpha$ at the transition and the speed and width of the transition.  If for any of the momentum values in Eq.~\eqref{eq:rangek} the causality bound from Eq.~\eqref{eq:causal} is violated, we deemed the region in parameter space for the corresponding $\{A_{c_s}, A_\alpha,\sigma, \wid\}$ causal.

%%%%%%%%%%%%%%
\section{Strong Coupling bounds} \label{app:sc}
In this appendix, we look at the strong coupling bounds in the EFT of inflation and also in the particular case we are considering in \eqref{eftofi}. To start the discussion let us consider action consider the following  cubic action for the Goldstone boson, 
\begin{align}
S_\pi=\int d^4x \sqrt{-g}\left[-\frac{M_{\mathrm{Pl}}^2\dot{H}}{c_s^2}\left(\dot\pi^2-c_s^2\frac{(\partial_i\pi)^2}{a^2}\right)+M_{\mathrm{Pl}}^2\dot H(1-c_s^{-2})\left(\dot\pi^3-\dot\pi\frac{(\partial\pi)^2}{a^2}\right)-\frac{4}{3}M_3^4\dot\pi^3\right] \ ,
\end{align}
We are interested in the limit where the weak coupling action stops being valid. It is useful to canonically normalise the fields to $\pi_c^2=2M_{\mathrm{Pl}}^2\vert\dot{H}\vert c_s^{-2}\pi^2$ and  define,
\begin{align}
  M_*^4\equiv M_{\mathrm{pl}}^2\vert\dot{H}\vert c_s^{-2}(1-c_s^2)^{-1} ,\qquad\mathrm{and}\qquad \frac{A}{c_s^2}\equiv-1+\frac{4}{3}\frac{M_3^4}{M_{\mathrm{pl}}^2\vert\dot{H}\vert}\frac{c_s^2}{c_s^2-1} ,
\end{align}
such that the action becomes,
\begin{align}
\mathcal{L}=-\frac{1}{2}(\dot\pi_c^2-c_s^2(\partial_i\pi_c)^2)+\frac{1}{M_\star^2}\left(\dot\pi_c(\partial_i\pi_c)^2 +A\dot\pi_c^3\right).
\end{align}
where we require $A\sim O(1)$ to not have an unnatural hierarchy between the two operators $\dot\pi_c^3$ and $(\partial_i\pi_c)^2\dot\pi_c)$~\cite{Senatore:2010jy}.
To translate $M_\star$ into an energy scale, it is necessary to carefully account for the units. Since the system is non-relativistic, the dispersion relation $\omega^2 = c_s^2 k^2$ must be used to relate energy ($\omega$) to momentum ($k$). Dimensional analysis shows that the canonical Goldstone field $\pi_c$ has units $$[\pi_c] = [k]^{3/2}[\omega]^{-1/2} \ . $$ From this, the interaction scale $M_\star^4$ acquires the units $[M_\star^4] = [k]^7[\omega]^{-3}$.
This implies that the strong coupling scale $\Lambda_\star$, defined in terms of energy, is related to $M_\star$ by

\begin{align}
\Lambda_\star^4 = M_\star^4 \times c_s^7 \ ,
\end{align}
or equivalently
\begin{align}
\Lambda_\star^4 \equiv M_{\mathrm{Pl}}^2 |\dot{H}| c_s^{-5}(1-c_s^2)^{-1} \ .
\end{align}

This result matches the estimate derived from the breakdown of perturbative unitarity, differing only by a factor of $4\pi$. Notice that these bounds are obtained by assuming the EFT satisfies the perturbative unitarity constraints for flat spacetime \cite{Cheung:2007st,Baumann:2011su}.

Let us consider now the case when $c_s=0$ and $\alpha$ is constant. We have that the action is given by,
\begin{align}
    S_\pi^{(2)}=\int d^4 x a^3 (2M_2^{4}-M_{\mathrm{Pl}}^2\dot H)\left[\dot\pi-\alpha\frac{(\partial_i^2\pi)}{H^2 a^4}\right]=\frac{1}{2}\int d^4 x a^3\left[\dot\pi_c^2-\alpha\frac{(\partial_i^2\pi_c)^2}{H^2a^3}\right] \ ,
\end{align}
where $\pi_c=\sqrt{2(M_2^4-2M_{\mathrm{Pl}}^2\dot H)}\pi$ and $\alpha=\frac{\bar{M}_2^2H^2}{2(2M_2^4-M_{\mathrm{Pl}}^2)}$. If we consider the interaction given by
\begin{align}
    S^{(3)}=\int d^4x a^3 M_2^4 \dot\pi\frac{(\partial_i\pi)^2}{a^2}=\int d^4x a^3 \frac{M_2^4}{(4M_2^4-2M_{\mathrm{Pl}}^2\dot H)^{3/2}} \dot\pi_c\frac{(\partial_i\pi_c)^2}{a^2} \ ,
\end{align}
We have that $M_\star^4=\frac{(4M_2^4-2M_{\mathrm{Pl}}^2\dot H)^{3}}{M_2^8}$. Again this has units of $[k]^7[\omega]^{-3}$. Using that the dispersion relation is $\omega^2=\frac{\alpha k^4}{H^2}$ we then get the energy scale,
\begin{align}
    \Lambda_\star\sim\alpha^{7/2} \frac{M_*^8}{H^7}\sim (M_2^4-2M_{\mathrm{Pl}}^2\dot H)^{5/2} \frac{\bar{M}_2^7}{M_2^{16}} \ .
\end{align}
Note that numerical factors of order $\pi$ can be obtained by looking at when perturbative unitarity breaks down.
Usually, $c_s$ very low can be achieved by assuming that  $M_2^4\gg M_{\mathrm{Pl}}^2\vert\dot H\vert$. This leads to 
\begin{align}
    \Lambda_\star\sim \frac{\bar{M}_2^7}{M_2^6} \ ,
    \label{LUV1}
\end{align}
in agreement with \cite{Baumann:2011su}. A rougher estimate can also be obtained by computing the correction to the two-point function from the interaction. In this case, we have that,
\begin{align}
    \frac{1}{M_2^4}\frac{\dot\pi_c^2}{8(2\pi)^4} \int d\omega d^3k \frac{k^4}{\left(\omega^2-\frac{\alpha k^4}{H^2}\right)}\sim \dot\pi_c^2\frac{\alpha^{-7/2}}{M_2^4 H^7}\frac{\sqrt{\omega} }{8(2\pi)^3} \ ,
\end{align}
which give the same energy scale that \eqref{LUV1} up to factors of $(2\pi)$:
\begin{align}
    \Lambda_\star\sim (2\pi)^3 \frac{\bar{M}_2^7}{M_2^6} \ ,
\end{align}
and it also controls the size of the loop corrections to the kinetic term.

\noindent\textbf{WBG action} If we now go back to the action we were considering in \eqref{eftofi}, the third order interactions are given~ by~\cite{Pirtskhalava:2015zwa},
\begin{align}
    S_\pi^{(2)}=\int d^4x a^3\left[-\left(2M_2^4+\frac{3}{2}\bar{M}_2^2\dot H-\frac{1}{2}\partial_t (\hat{M}_1^3)\right)\dot\pi\frac{(\partial_i\pi)^2}{a^2}+\frac{\hat{M}_1^3}{2a^4}(\partial_i^2\pi)(\partial_j \pi)^2+\bar{M}_2^2\dot\pi\frac{(\partial_i^2\pi)^2}{a^4}\right],
\end{align}
The first term is the same we already considered in \eqref{LUV1}. Note that due to the WBG symmetry \eqref{eq:WBG} the coefficient simplifies to $M_2^4\dot\pi(\partial_i\pi)^2/a^2$, implying the same energy scales we considered before. However, as we will see, in this case, this is not the strong coupling scale.  If we canonically normalise the action we find, 
\begin{align}
    S_\pi^{(3)}&\supset \int d^4xa^3\left[-\frac{1}{M_2^2}\dot\pi_c\frac{(\partial_i\pi_c)^2}{a^2}+\frac{\hat{M}_1^3}{(8M_2^6)}\frac{(\partial_i^2\pi_c)(\partial^j\pi_c)^2}{2a^4}+\frac{\bar{M}_2^2}{(8M_2^6)}\dot\pi_c \frac{(\partial_i^2\pi_c)^2}{a^4} \right] \ ,\nonumber\\
    &\sim \int d^4 xa^3 \left[\frac{1}{M_2^2}\dot\pi_c\frac{(\partial_i\pi_c)^2}{a^2}+\frac{1}{(M_3^*)^3}\frac{(\partial_i^2\pi_c)(\partial^j\pi_c)^2}{2a^4}+\frac{1}{(M_4^*)^4}\dot\pi_c \frac{(\partial_i^2\pi_c)^2}{a^4}\right] \ ,
\end{align}
where we have labelled the mass scales according to their order in the action. 
By dimensional analysis, we can estimate how the mass scales of each operator are related to energy scales. Assuming that $\alpha$ dominates we find for each operator that, 
\begin{align}
\Lambda_2&\sim \frac{\bar{M}_2^7}{M_2^6}\sim \frac{H}{\alpha^{5/2}} \ ,\\
    \Lambda_3&\sim \frac{\bar{M}_2^7M_2^2}{\hat{M}_1^8}\sim \frac{H}{\sqrt{\alpha}} \ ,\\
    \Lambda_4&\sim \frac{\bar{M}_2^3}{M_2^2}\sim \frac{\Lambda_3^3}{H^2}\sim\frac{\Lambda_3}{\sqrt{\alpha}} \ ,
\end{align}
where for $\Lambda_3$ we have used that the EFT requires that  $M_2^4\sim \hat{M}_1^3 H$. Since we can write that $\Lambda_2\sim \Lambda_3\alpha^{-2}$, it becomes clear that the first operator to become strongly coupled is $\frac{1}{(M_3^*)^3a^4}\partial_i^2\pi_c (\partial^j\pi_c)^2$, hence we have that when $\alpha$ dominates $\Lambda_\star=\Lambda_3$.
\bibliographystyle{utphys}
\bibliography{biblio.bib}
\end{document}